\documentclass[
    reprint, twocolumn,
    aps,prb,10pt,amsmath,amssymb,
    longbibliography,superscriptaddress,
    showpacs,preprintnumbers,
    groupedaddress]{revtex4-1}

\usepackage[colorlinks=true,citecolor=blue,linkcolor=blue]{hyperref}
\usepackage{oubraces}
\usepackage{bm,bbm,mathbbol}
\usepackage{physics}
\usepackage{mathtools}
\usepackage{amsmath, amsthm, amssymb, amsfonts}
\usepackage{bbold}
\usepackage{esvect}

\usepackage{epsfig}
\usepackage{array}
\usepackage{multirow}
\usepackage{graphicx}
\usepackage[normalem]{ulem}

\usepackage{float}
\usepackage[section]{placeins}
\usepackage{afterpage}

\usepackage{lipsum}

\usepackage{xifthen}

\makeatletter
\newcommand\footnoteref[1]{\protected@xdef\@thefnmark{\ref{#1}}\@footnotemark}
\makeatother

\newcommand{\nnnl}{\nonumber \\}

\newcommand{\qe}[0]{{\sc Quantum ESPRESSO}}
\newcommand{\wannier}[0]{{\sc Wannier90}}

\newcommand{\myeqref}[1]{Eq.~\eqref{#1}}

\newcommand{\mb}[1]{{\boldsymbol{\mathbf{#1}}}}
\newcommand{\nk}[0]{{n\mathbf{k}}}
\newcommand{\mk}[0]{{m\mathbf{k}}}

\newcommand{\veps}[0]{\varepsilon}

\newcommand{\opH}[0]{\hat{H}}
\newcommand{\opV}[0]{\hat{V}}
\newcommand{\opc}[0]{\hat{c}}
\newcommand{\opcd}[0]{\hat{c}^\dagger}
\newcommand{\subii}[0]{{\mathrm{(ii)}}}
\newcommand{\subei}[0]{{\mathrm{(ei)}}}
\newcommand{\subie}[0]{{\mathrm{(ie)}}}
\newcommand{\subee}[0]{{\mathrm{(ee)}}}
\newcommand{\der}[1]{\partial^{{#1}}}
\newcommand{\abc}[0]{{a;bc}}
\newcommand{\acb}[0]{{a;cb}}
\newcommand{\sabc}[0]{{s,a;bc}}
\newcommand{\omegaall}[0]{{(\omega;\omega_1,\omega_2)}}

\newcommand{\mcM}[0]{{\mathcal{M}}}

\newcommand{\mcE}[0]{{\mathcal{E}}}

\newcommand{\opmcM}[0]{{\hat{\mathcal{M}}}}
\newcommand{\opmcT}[0]{{\hat{\mathcal{T}}}}
\newcommand{\opmcP}[0]{{\hat{\mathcal{P}}}}
\newcommand{\opmcPT}[0]{{\hat{\mathcal{P}}\hat{\mathcal{T}}}}

\newcommand{\Rtilde}[0]{{\widetilde{R}}}

\begin{document}

\title{A comprehensive theory of second-order spin photocurrents}

\author{Jae-Mo Lihm}
\email{jaemo.lihm@gmail.com}
\author{Cheol-Hwan Park}
\email{cheolhwan@snu.ac.kr}
\affiliation{Center for Correlated Electron Systems, Institute for Basic Science, Seoul 08826, Korea}
\affiliation{Department of Physics and Astronomy, Seoul National University, Seoul 08826, Korea}
\affiliation{Center for Theoretical Physics, Seoul National University, Seoul 08826, Korea}

\date{\today}

\begin{abstract}
The spin photocurrents, direct currents induced by light, hold great promise for introducing new elements to spintronics.
However, a general theory for spin photocurrents in real materials which is applicable to systems with spin-orbit coupling or noncollinear magnetism is absent.
Here, we develop such a general theory of second-order spin photocurrents.
We find that the second-order spin photocurrents can be classified into Drude, Berry curvature dipole, shift, injection, and rectification currents, which have different physical origins and symmetry properties.
Surprisingly, our theory predicts a direct pure spin rectification current in an insulator induced by photons with energies lower than the material band gap.
This phenomenon is absent in the case of the charge photocurrent.
We find that the pure spin current of BiTeI induced by subgap light is large enough to be observable in experiments.
Moreover, the subgap pure spin photocurrent is highly tunable with the polarization of light and the flowing direction of the spin photocurrent.
This study lays the groundwork for the study of nonlinear spin photocurrents in real materials and provides a route to engineer light-controlled spin currents.
\end{abstract}

\maketitle

%=========================================================
\section{Introduction}
Photocurrents, the current response of materials under light irradiation, 
has potential in solar cell applications~\cite{2013Grinberg}
and optoelectronic devices~\cite{2010BonaccorsoOptoelec}.
The spin version of the photocurrent response is gaining much attention as a promising tool in spintronics~\cite{2005BhatSpinInjection,2008Ivchenko,2013YoungSpin}.
Spin photocurrents enable an optical generation of spin-polarized currents, essential building blocks of spintronics.
Spin photocurrents have been intensely investigated, both theoretically~\cite{2005BhatSpinInjection,2005Tarasenko,2013YoungSpin} and experimentally~\cite{2005Zhao,2006GanichevSpin}.

However, a general theory of bulk spin photocurrents is yet to be developed, despite its importance and the interest in it.
Most of the existing studies either deal with spin-collinear systems, where the spin-up and spin-down electrons are completely decoupled~\cite{2013YoungSpin,2020FeiSpin,2020XiaoSpinCollinear},
or consider a specific part of the full response, such as the Drude current~\cite{2001GanichevSpin,2017HamamotoSpin} or the injection current~\cite{2005BhatSpinInjection,2007CuiSpinInjection,2021FeiSpinInjection}.
The spin shift current in spin-orbit coupled systems has been investigated in a few studies~\cite{2005BhatSpinInjection,2017KimSpinshift,2020XuSpin,2020MuSpinPhotocurrent} although we show that those descriptions for the spin shift current are not complete.
In addition, noncollinear magnetism enables the generation of spin-polarized currents~\cite{2017ZeleznyNoncollinear} and spin Hall currents~\cite{2018ZhangSHE} without spin-orbit coupling.
However, current theories cannot describe spin photocurrents in such systems.
This situation is in sharp contrast to the case of charge photocurrents~\cite{2000Sipe}, where a complete classification of second-order responses was recently developed~\cite{2019ZhangPhotocurrent,2020DejuanPhotocurrent,2020Holder,2020AhnBPVE,2020GaoIntrinsicFS,2021WatanabePhotocurrent}.
Thus, a complete, generally applicable theory of spin photocurrent at the level of its charge counterpart is highly desirable.

In this paper, we develop a complete theory of second-order spin photocurrents, which describes the spin and charge photocurrents in a unified framework.
Using perturbation theory in the length gauge, we derive the expression for the second-order spin conductivity tensor.
We show that the spin photocurrent can be classified into Drude, Berry curvature dipole, shift, injection, and rectification currents.
We find that the spin shift and rectification currents can be written in terms of the complex-valued spin shift vector, a gauge-invariant quantity proposed in this work.
Interestingly, the spin rectification current is nonzero in insulators, in sharp contrast with its charge counterpart, the ``intrinsic Fermi surface'' current, which is always zero in insulators~\cite{2020DejuanPhotocurrent,2020GaoIntrinsicFS,2021WatanabePhotocurrent}.
Most importantly, finite spin rectification currents can be generated even by light whose frequency is lower than the band gap.
These subgap pure spin currents are highly tunable with the polarization of light.
We demonstrate our theory and findings by presenting our calculations on the spin and charge photoconductivity of BiTeI.

%=========================================================
\section{Theory}
\subsection{Second-order spin photoconductivity}
We study the response of the system to external electric fields using the Schr\"odinger equation for the one-particle reduced density matrix~\cite{2017VenturaGauge}.
We write the responses in terms of the velocity matrix element
\begin{equation}
    v^a_{mn,\mb{k}} = \mel{u_\mk}{\partial^a H_0(\mb{k})}{u_\nk} / \hbar,
\end{equation}
occupation factor $f_\mk = 1 / \{1 + \exp[(\veps_\mk - \mu) / k_{\rm B} T]\}$, and frequency $\omega_{mn} = (\veps_m - \veps_n) / \hbar$.
Here, $H_0(\mb{k})$ is the periodic Bloch Hamiltonian, $\ket{u_\mk}$ the periodic part of the Bloch wavefunction of state $m$ with crystal momentum $\mb{k}$, $\veps_\mk$ the band energy, $T$ the temperature, $\mu$ the chemical potential, and $\partial^a = \partial/\partial k_a$.
In the following, we omit the subscript $\mb{k}$ for brevity.
We also define $f_{mn} = f_m - f_n$.
We define the Berry connection $\xi^a_{mn} = i \braket{u_m}{\partial^a u_n}$, which satisfies
\begin{equation} \label{eq:xi_and_v}
    \xi^a_{mn} = -i v^a_{mn} / \omega_{mn}
\end{equation}
for non-degenerate states with $m \neq n$.

The spin current is the expectation value of the spin-current operator, whose matrix element is
\begin{equation} \label{eq:j_def}
    j^{s,a}_{mn} = \frac{1}{2}\mel{u_m} {\acomm{S^s}{v^a}} {u_n},
\end{equation}
where $S^s$ is the spin operator with $s=x,y,z$.
We also let $j^{0,a} = v^a$ be the charge-current operator by defining $S^0$ to be the identity operator.
For later use, we define
\begin{equation} \label{eq:Deltaj_def}
 \Delta j^{s,a}_{mn} = j^{s,a}_{mm} - j^{s,a}_{nn}.
\end{equation}
(See Appendix~\ref{sec:app_spin_current} for the discussion on alternate forms of the spin-current operator.)

Another key quantity for nonlinear spin currents is the {\it spin-velocity derivative} of the velocity operator, which we define as follows:
\begin{equation} \label{eq:sder_def}
    d^{s,b;a}_{mn}
    = j^{s,ab}_{mn}
    + \sum_{p\neq m} \frac{j^{s,a}_{mp} v^b_{pn}}{\omega_{mp}}
    + \sum_{p\neq n} \frac{v^b_{mp} j^{s,a}_{pn}}{\omega_{np}}.
\end{equation}
Here, we also defined
\begin{equation}
    j^{s,ab}_{mn} = \frac{1}{2} \mel{u_m}{\acomm{S^s}{ \der{a}\der{b}H_0(\mb{k})}}{u_n}.
\end{equation}
The name ``spin-velocity derivative'' reflects the fact that for the charge current, $d^{0,b;a}_{mn}$ is identical to the generalized derivative of the velocity operator:
\begin{equation} \label{eq:sder_charge}
    d^{s=0,b;a}_{mn}
    = \der{a} v^b_{mn} - i(\xi^a_{mm} - \xi^a_{nn}) v^b_{mn}.
\end{equation}
For spin-collinear systems, the spin-velocity derivatives are equivalent to the charge counterparts except for a minus sign for the spin-down states.
On the contrary, for spin-noncollinear systems, the spin-velocity derivatives cannot be simply related to the charge counterparts.

In this work, we focus on the second-order response, which is the lowest order where a DC current response can occur under AC driving fields.
We consider external fields with frequencies $-\Omega$ and $\Omega + \omega$ and study the second-order response with frequency $\omega$ in the DC limit $\omega \rightarrow 0$ with $\Omega$ fixed.
We assume a clean system with an infinite carrier lifetime as done in previous studies of charge currents~\cite{2000Sipe,2019ParkerDiagram,2019ZhangPhotocurrent,2020DejuanPhotocurrent,2020GaoIntrinsicFS,2020AhnBPVE,2021WatanabePhotocurrent}.
Thus, the DC limit should not be taken literally: the photocurrent should be understood as that from a difference frequency generation where $\Omega \gg \omega \gg 1/\tau$~\cite{2020DejuanPhotocurrent}, where $\tau$ is the characteristic lifetime of the bands.

Watanabe and Yanase~\cite{2021WatanabePhotocurrent} applied the general results of Ref.~\cite{2017VenturaGauge} to the second-order responses of charge currents.
In this work, we extend this formalism~\cite{2017VenturaGauge,2021WatanabePhotocurrent} to spin currents.
The second-order spin and charge photoconductivity is the sum of five distinct responses: Drude, Berry curvature dipole (BCD), shift, injection, and rectification photoconductivities.
The formula for each photoconductivity is as follows.
\begin{align} \label{eq:sigma_drude}
    \sigma^\sabc_\mathrm{Drude}(\Omega)
    = \frac{q^3}{2\hbar^2 V \Omega^2} \sum_{\mb{k},m} j^{s,a}_{mm} \der{b} \der{c} f_m
\end{align}
\begin{align} \label{eq:sigma_BCD}
    \sigma^\sabc_\mathrm{BCD}(\Omega)
    = \frac{-i q^3}{\hbar^2 V \Omega} \sum_{\substack{\mb{k},m,n \\ m \neq n}} \frac{\Im (j^{s,a}_{mn} v^b_{nm})}{\omega_{mn}^2} \der{c} f_{m}
    - (b \leftrightarrow c)
\end{align}
\begin{align} \label{eq:sigma_inj}
    \sigma^\sabc_\mathrm{inj.,\tau}(\Omega)
    &= \tau \eta^\sabc_\mathrm{inj.}(\Omega) \\
    &= -\tau\frac{\pi q^3}{\hbar^2 \Omega^2 V} \sum_{\mb{k},m,n} f_{mn} \Delta j^{s,a}_{mn} v^b_{mn} v^c_{nm} \delta(\Omega + \omega_{mn}) \nonumber
\end{align}
\begin{align} \label{eq:sigma_shift}
    \sigma^\sabc_\mathrm{shift}(\Omega)
    = \frac{i \pi q^3}{2\hbar^2 V\Omega^2} \sum_{\mb{k},m,n}
    &(d^{s,b;a}_{mn} v^c_{nm} - d^{s,c;a}_{nm} v^b_{mn}) \nnnl
    &\times f_{mn} \delta(\Omega + \omega_{mn})
\end{align}
\begin{align} \label{eq:sigma_rec}
    &\sigma^\sabc_\mathrm{rect.}(\Omega)
    = \frac{q^3}{2 \hbar^2 V} \sum_{\substack{\mb{k},m,n \\ m\neq n}} \frac{f_{mn}}{\omega_{mn}^2}
    \Bigg[ \Big( 
    d^{s,b;a}_{mn} v^c_{nm} + d^{s,c;a}_{nm} v^b_{mn} \nnnl
    &- \frac{2\Delta j^{s,a}_{mn} v^b_{mn}  v^c_{nm}}{\omega_{mn}} 
    \Big) \mathrm{P} \frac{1}{\Omega + \omega_{mn}}
    - \frac{\Delta j^{s,a}_{mn} v^b_{mn} v^c_{nm}}{(\Omega + \omega_{mn})^2} \Bigg]
\end{align}
Here, $V$ is the volume of the system, and $q$ the charge of an electron.
The detailed derivation of these equations is given in Appendix~\ref{sec:app_derivation}.
For the injection current, we introduced a phenomenological relaxation time $\tau$~\cite{2018PassosVelocitygauge}.
This treatment is needed to avoid divergence in the DC limit and is justified by calculations based on the Floquet formalism~\cite{2017deJuanCPVE}.

The Drude~[\myeqref{eq:sigma_drude}] and BCD~[\myeqref{eq:sigma_BCD}] terms contain derivatives of the occupation factor.
Therefore, the corresponding currents are zero in insulators.
While the shift and injection currents originate from the absorptive (resonant) response, the rectification current originates from the reactive (non-resonant) response.
The injection and shift currents are nonzero only when occupied and unoccupied states are resonantly coupled by light: $\Omega + \omega_{mn} = 0$.
In contrast, the rectification current does not require an energy-conserving transition.

The charge rectification current was discovered only recently~\cite{2020DejuanPhotocurrent,2020GaoIntrinsicFS,2021WatanabePhotocurrent}.
It was termed the ``intrinsic Fermi surface'' contribution because one can convert \myeqref{eq:sigma_rec} for the charge current case into a Fermi-surface integral using \myeqref{eq:sder_charge}:
\begin{align} \label{eq:sigma_rec_charge}
    &\sigma^{i=0,\abc}_\mathrm{rect.}
    = \frac{q^3}{2 \hbar^2 V} \sum_{\mb{k},m,n}
    f_{mn} \der{a} \left( \frac{v^{b}_{mn} v^c_{nm}}{\omega_{mn}^2} \mathrm{P} \frac{1}{\Omega + \omega_{mn}} \right) \nnnl
    &= -\frac{q^3}{2 \hbar^2 V} \sum_{\mb{k},m,n}
    (\der{a} f_{mn}) \frac{v^{b}_{mn} v^c_{nm}}{\omega_{mn}^2} \mathrm{P} \frac{1}{\Omega + \omega_{mn}}.
\end{align}
Hence, the charge rectification current is zero in insulators.

However, we find that the spin rectification current is nonzero even in insulators.
One of the reasons for the contrast with the charge case is that the spin-velocity derivative $d^{s,b;a}_{mn}$ with $s\neq0$ is not an actual derivative with respect to the crystal momentum.
In other words, the spin-velocity derivative, like the spin Berry curvature~\cite{2008GuoSHE}, is not a geometric quantity.
Thus, in the spin current case, one cannot rewrite $\sigma^{s,\abc}_\mathrm{rect.}$ as a Fermi-surface integral.

Since the rectification current does not require a resonant coupling of occupied and unoccupied states, it leads to a subgap spin photocurrent: a DC spin photocurrent under irradiation of light with a frequency less than the band gap of an insulating system.
Such a subgap response is unique to the spin photocurrent, while absent in the charge counterpart.

We note that the previously reported equations for the spin shift current~\cite{2017KimSpinshift,2020XuSpin,2020MuSpinPhotocurrent} are not equivalent to our results [Eqs.~(\ref{eq:sigma_shift}, \ref{eq:sigma_shift_with_R})].
Reference \cite{2017KimSpinshift}, which uses the velocity gauge, considers only the diamagnetic two-photon output vertex and ignores the contribution of a diagram with three one-photon vertices.
The photoconductivity formula of Refs.~\cite{2020XuSpin,2020MuSpinPhotocurrent}, which is also based on the velocity gauge, does not take the two-photon vertex into account.
To obtain the correct result using the velocity gauge formalism, one needs to consider all diagrams consisted of one-, two-, and three-photon vertices and apply related sum rules~\cite{2018PassosVelocitygauge,2019ParkerDiagram,2020GaoIntrinsicFS}.

%========================================================
\subsection{Writing photoconductiviy in terms of the complex shift vector}

To understand the physical mechanism behind the spin shift and rectification currents, we now show that these currents can be written in terms of the ``spin shift vector'' which is a gauge-invariant quantity we introduce in this work.

For incident light with polarization vector $\mb{\mcE}$, the second-order DC current response reads
\begin{equation} \label{eq:sigma_bc_to_E}
    \sigma^{s,a;\mcE}(\Omega) = \sum_{b, c} \sigma^{s,a;bc}(\Omega) \mcE^b \mcE^{*c}.
\end{equation}
We also define
\begin{equation} \label{eq:sigma_O_E}
    \mathcal{O}^\mcE_{mn} = \sum_a \mathcal{O}^a_{mn} \mcE^a
\end{equation}
for any vector or tensor operator $\mb{\mathcal{O}}$.
For a Hermitian matrix $\mathcal{O}$, the following holds:
\begin{equation} \label{eq:supp_O_E_Hermitian}
    \left( \mathcal{O}^{\mcE^*}_{nm} \right)^*
    = \sum_a \left( \mathcal{O}^a_{nm} \mcE^{*a} \right)^*
    = \sum_a \mathcal{O}^a_{mn} \mcE^{a}
    = \mathcal{O}^{\mcE}_{mn}.
\end{equation}

Following Ref.~\cite{2020AhnBPVE}, we define the complex shift vector between two different bands $m$ and $n$ for light polarization $\mb{\mcE}$ as
\begin{equation}
    \Rtilde^{\mcE;a}_{mn}
    = i \frac{\mathcal{D}^a \xi^\mcE_{mn}}{\xi^\mcE_{mn}}
    = \xi^a_{mm} - \xi^a_{nn} + i \partial^a \log \xi^\mcE_{mn}
\end{equation}
Note that we consider a fixed light polarization because the shift vector does not transform like a vector with respect to the rotation of the light polarization.
By using \myeqref{eq:xi_and_v}, the complex shift vector can be written in terms of the velocity matrix and its generalized derivative:
\begin{equation} \label{eq:complex_shift_in_v}
    \Rtilde^{\mcE;a}_{mn} = i\left( \frac{d^{0,\mcE;a}_{mn}}{v^\mcE_{mn}} - \frac{\Delta v^a_{mn}}{\omega_{mn}} \right).
\end{equation}
The standard shift vector~\cite{2000Sipe} is the real-part of the complex shift vector:
\begin{equation} \label{eq:shift_vector}
    R^{\mcE;a}_{mn}
    = \xi^a_{mm} - \xi^a_{nn} - \partial^a \arg \xi^\mcE_{mn}
    = \Re \Rtilde^{\mcE;a}_{mn},
\end{equation}
where `$\arg$' denotes the argument of a complex number.
This real-valued shift vector describes the change in the position of the electron in the transition from band $n$ to $m$.

Now, we define the spin shift vector as the spin generalization of \myeqref{eq:complex_shift_in_v}:
\begin{equation} \label{eq:spin_shift_vector}
    \Rtilde^{s,\mcE;a}_{mn} = i\left( \frac{d^{s,\mcE;a}_{mn}}{v^\mcE_{mn}} - \frac{\Delta j^{s,a}_{mn}}{\omega_{mn}} \right).
\end{equation}
The spin shift vector is invariant to the gauge transformation $\ket{u_\mk} \rightarrow \ket{u_\mk}e^{i\phi_\mk}$ and thus represents a physical quantity.
By an analogy to the charge case, the real part of the spin shift vector can be interpreted as the shift in the spin position in the transition from band $n$ to $m$.
However, in contrast to the charge case, it is not possible to write $\Rtilde^{s,\mcE;a}_{mn}$ directly in terms of the matrix elements between bands $m$ and $n$ and their derivatives due to the nongeometric nature of the spin-current operator.

To rewrite the shift photoconductivity [\myeqref{eq:sigma_shift}] and the rectification conductivity [\myeqref{eq:sigma_rec}] in terms of the spin shift vector, we use the following property:
\begin{align} \label{eq:supp_Rtilde_Hermitian}
    \left( \Rtilde^{s,\mcE;a}_{mn} \right)^*
    =& -i\left( \frac{d^{s,\mcE;a}_{mn}}{v^{\mcE}_{mn}} - \frac{\Delta j^{s,a}_{mn}}{\omega_{mn}} \right)^* \nnnl
    =& -i\left( \frac{d^{s,\mcE^*;a}_{nm}}{v^{\mcE^*}_{nm}} - \frac{\Delta j^{s,a}_{nm}}{\omega_{nm}} \right) \nnnl
    =& -\Rtilde^{s,\mcE^*;a}_{nm}.
\end{align}

Then, by substituting \myeqref{eq:sigma_shift} into \myeqref{eq:sigma_bc_to_E}, we can write the spin shift current for light polarization $\mcE$ in terms of the spin shift vector as
\begin{widetext}
\begin{align} \label{eq:sigma_shift_with_R}
    \sigma^{s,a;\mcE}_\mathrm{shift}(\Omega)
    =& \frac{i \pi q^3}{2\hbar^2 V\Omega^2} \sum_{\mb{k},m,n}
    (d^{s,\mcE;a}_{mn} v^{\mcE^*}_{nm} - d^{s,\mcE^*;a}_{nm} v^\mcE_{mn})
    f_{mn} \delta(\Omega + \omega_{mn}) \nnnl
    =& \frac{i \pi q^3}{2\hbar^2 V\Omega^2} \sum_{\mb{k},m,n}
    (-i \Rtilde^{s,\mcE;a}_{mn} v^\mcE_{mn} v^{\mcE^*}_{nm}
    + \Delta j^{s,a}_{mn} \frac{v^\mcE_{mn} v^{\mcE^*}_{nm}}{\omega_{mn}}
    + i \Rtilde^{s,\mcE^*;a}_{nm} v^{\mcE^*}_{nm} v^\mcE_{mn}
    - \Delta j^{s,a}_{nm} \frac{v^{\mcE^*}_{nm} v^\mcE_{mn}}{\omega_{nm}})
    f_{mn} \delta(\Omega + \omega_{mn}) \nnnl
    =& \frac{i\pi q^3}{2\hbar^2 V\Omega^2} \sum_{\mb{k},m,n}
    (-i \Rtilde^{s,\mcE;a}_{mn} + i \Rtilde^{s,\mcE^*;a}_{nm})
    f_{mn} \omega_{mn}^2 \xi^\mcE_{mn} \xi^{\mcE^*}_{nm} \delta(\Omega + \omega_{mn}) \nnnl
    =& \frac{\pi q^3}{\hbar^2 V} \sum_{\mb{k},m,n}
    \Re \left(\Rtilde^{s,\mcE;a}_{mn} \right)
    f_{mn} \abs{\xi^\mcE_{mn}}^2 \delta(\Omega + \omega_{mn}).
\end{align}
\end{widetext}
In the last equality, we used \myeqref{eq:supp_Rtilde_Hermitian}.
For the charge case $s=0$, \myeqref{eq:sigma_shift_with_R} agrees with the known formula for the shift current under linear and circular polarization~\cite{2000Sipe,2020AhnBPVE,2021WatanabePhotocurrent}.
(Note that according to our derivation, the shift current due to circularly polarized light corresponds to the ``gyration current'' term in Ref.~\cite{2021WatanabePhotocurrent}.)
The novel finding of this work is that the spin shift current can be written in the same form by defining the spin shift vector as \myeqref{eq:spin_shift_vector}.

The last line of \myeqref{eq:sigma_shift_with_R} has the form of the Fermi golden rule.
Hence, the spin shift current can be interpreted as the current due to the spin density dipole generated by the transition from state $n$ to $m$.
This interpretation parallels the case of charge shift current~\cite{1981volBaltzShift,2000Sipe}.
However, in contrast to the charge case~[\myeqref{eq:shift_vector}], it is not possible to write the spin shift vector $\Rtilde^{s,\mcE;a}_{mn}$ only in terms of the low-energy wavefunctions due to the nongeometric nature of the spin-current operator.

For the spin rectification current, we find
\begin{widetext}
\begin{align} \label{eq:sigma_rec_with_R}
    \sigma^{s,a;\mcE}_\mathrm{rect.}(\Omega)
    =& \frac{q^3}{2 \hbar^2 V} \sum_{\substack{\mb{k},m,n \\ m\neq n}} \frac{f_{mn}}{\omega_{mn}^2}
    \Bigg[ \Big( 
    d^{s,\mcE;a}_{mn} v^{\mcE^*}_{nm} + d^{s,\mcE^*;a}_{nm} v^\mcE_{mn}
    - \frac{2\Delta j^{s,a}_{mn} v^\mcE_{mn}  v^{\mcE^*}_{nm}}{\omega_{mn}} 
    \Big) \mathrm{P} \frac{1}{\Omega + \omega_{mn}}
    - \frac{\Delta j^{s,a}_{mn} v^\mcE_{mn} v^{\mcE^*}_{nm}}{(\Omega + \omega_{mn})^2} \Bigg] \nnnl
    =& \frac{q^3}{2 \hbar^2 V} \sum_{\substack{\mb{k},m,n \\ m\neq n}} \frac{f_{mn}}{\omega_{mn}^2}
    \Bigg[ \left( -i \Rtilde^{s,\mcE;a}_{mn} v^\mcE_{mn} v^{\mcE^*}_{nm} -i \Rtilde^{s,\mcE^*;a}_{nm} v^{\mcE^*}_{nm} v^\mcE_{mn} \right) \mathrm{P} \frac{1}{\Omega + \omega_{mn}}
    - \frac{\Delta j^{s,a}_{mn} v^\mcE_{mn} v^{\mcE^*}_{nm}}{(\Omega + \omega_{mn})^2} \Bigg] \nnnl
    =& \frac{q^3}{\hbar^2 V} \sum_{\substack{\mb{k},m,n \\ m\neq n}} f_{mn}
    \Bigg[ \Im \left( \Rtilde^{s,\mcE;a}_{mn} \right) \mathrm{P} \frac{1}{\Omega + \omega_{mn}}
    - \frac{\Delta j^{s,a}_{mn}}{2(\Omega + \omega_{mn})^2} \Bigg] \abs{\xi^\mcE_{mn}}^2.
\end{align}
\end{widetext}
In the last equality, we used \myeqref{eq:supp_Rtilde_Hermitian}.

Interestingly, the shift and rectification currents are proportional to the real and imaginary parts of the shift vector, respectively.
While the real part of the shift vector has been understood as the positional shift of the electron wave packet~\cite{1981volBaltzShift,2000Sipe}, we find here that the imaginary part plays an important role in the charge and spin rectification currents.
The physical meaning of the imaginary part remains a subject of future study.

\subsection{Symmetry of photoconductivity}
\begin{table}[]
\centering
\begin{tabular}{|c|c|c|c|}
\hline
    \multicolumn{1}{|c|}{\multirow{2}{*}{}} & \multicolumn{1}{c|}{\multirow{2}{*}{Current type}} & \multicolumn{2}{c|}{Jahn symbols} \\ \cline{3-4}
    \multicolumn{1}{|c|}{} & \multicolumn{1}{c|}{} & Linear pol. & Curcular pol.         \\ \hline
    \multirow{4}{*}{\begin{tabular}[c]{@{}c@{}}Charge\\$\sigma^{0,a;bc}$\end{tabular}}
    & Drude                    & a[V3]       & -           \\
    & Berry curvature dipole   & -           & V\{V2\}     \\
    & Shift                    & V[V2]       & aV\{V2\}    \\
    & Injection, Rectification & aV[V2]      & V\{V2\}     \\ \hline
    \multirow{4}{*}{\begin{tabular}[c]{@{}c@{}}Spin\\$\sigma^{s,a;bc}$\end{tabular}}
    & Drude                    & eVV[V2]     & -           \\
    & Berry curvature dipole   & -           & aeVV\{V2\}  \\
    & Shift                    & aeVV[V2]    & eVV\{V2\}   \\
    & Injection, Rectification & eVV[V2]     & aeVV\{V2\}  \\ \hline
\end{tabular}
\caption{
Jahn symbols for the second-order charge and spin photoconductivities in the clean limit. Hyphens indicate that the corresponding currents are always zero.}
\label{table:Jahn_symbol}
\end{table}

Next, we study the symmetry properties of the photocurrents.
To represent the symmetry properties, we use the Jahn symbols~\cite{1949Jahn,2019GallegoMTensor}.
Let us briefly explain the Jahn symbols.
Each `V' term corresponds to the vector indices.
The number of `V' corresponds to the tensor rank.
V's inside [\,] and \{\,\} denote symmetric and anti-symmetric indices, respectively.
Symbol `e' and `a' denotes that the tensor is axial and odd under time reversal, respectively.

The real symmetric (imaginary anti-symmetric) component of the photoconductivity tensor, $\Re(\sigma^\sabc + \sigma^{s,a;cb})$ ($\Im(\sigma^\sabc - \sigma^{s,a;cb})$) corresponds to the generation of a current under linearly (circularly) polarized light.
In Appendix~\ref{sec:supp_symmetry}, we derive the symmetry transformation properties of the five charge photoconductivity tensors using the symmetry of the velocity matrix elements.

One can easily derive the symmetry of the spin photoconductivity from the symmetry of the corresponding charge photoconductivity.
The second-order spin photoconductivity $\sigma^{s,a;bc}$ is a rank-4 tensor due to the additional vector describing spin polarization.
Equation~\eqref{eq:j_def} shows that the spin velocity transforms under symmetry operations like the product of the charge velocity and the spin polarization vector. The same relation holds for the spin and charge conductivity.
The spin polarization vector is a time-reversal odd, axial vector.
Thus, the Jahn symbol for the spin photoconductivity can be obtained by multiplying `aeV' to the Jahn symbol of the corresponding charge photoconductivity.
The only exception is the Drude current.
As explained in Appendix~\ref{sec:supp_symmetry_drude}, the charge Drude current has an additional symmetry for swapping the current and field direction indices [\myeqref{eq:supp_symm_drude_3}].
An analogous expression does not exist for the spin Drude current because the spin velocity $j^{s,a}_{mm}$ for $s=x,y,z$ is not a derivative of some quantity with respect to the crystal momenta.

In Table~\ref{table:Jahn_symbol}, we summarize the symmetry of the charge and spin photoconductivity tensors.
The nonzero, independent coefficients for the photoconductivity tensors of a given material can be easily identified using the MTensor program~\cite{2019GallegoMTensor}.

Let us focus on the role of the spatial inversion $\opmcP$, time reversal $\opmcT$, and space-time inversion $\opmcPT$.
The properties of the second-order charge conductivity in inversion- or time-reversal-symmetric systems were recently studied~\cite{2020AhnBPVE,2021WatanabePhotocurrent}.
Here, we extend the analysis to the spin conductivity.

Spatial inversion $\opmcP$ flips the sign of odd-rank vectors and gives an additional minus sign to axial tensors.
Since the charge photoconductivity is an odd-rank polar (non-axial) tensor, it obtains a minus sign under $\opmcP$.
The spin photoconductivity is an even-rank axial tensor, so it also obtains a minus sign under $\opmcP$.
Therefore, in an inversion-symmetric system, both the spin and charge second-order photoconductivity is always zero.

Time-reversal operation $\opmcT$ flips the sign of time-reversal-odd tensors, which have `a' in their Jahn symbols.
Therefore, among the second-order photoconductivities, only those {\it without} `a' in their Jahn symbols in Table~\ref{table:Jahn_symbol} are allowed in a $\opmcT$-symmetric system.
Therefore, in a $\opmcT$-symmetric system, the spin (charge) shift current is induced by circularly (linearly) polarized light, and the spin (charge) injection and rectification current is induced by linearly (circularly) polarized light.
Also, the spin Drude (charge BCD) current is induced by linearly (circularly) polarized light.

For a $\opmcPT$-symmetric system, the relation is the opposite: only the photocurrents with `a' in their Jahn symbols are allowed.
The reason is as follows.
All second-order spin and charge photoconductivities are all odd under $\opmcP$.
For the $\opmcPT$ symmetry operation, the $(-1)$ factor from $\opmcP$ is multiplied from the sign factor coming from $\opmcT$.
Thus, for both the spin and charge cases, if the photoconductivity is even (odd) under $\opmcT$, it is odd (even) under $\opmcPT$.
In a $\opmcPT$-symmetric system, only tensors that are even under $\opmcPT$ can be nonzero.
Therefore, among the second-order photoconductivities, only those {\it with} `a' in their Jahn symbols (i.e., those which were not allowed in $\opmcT$-symmetric system systems) are allowed in a $\opmcPT$-symmetric system.

The injection and rectification currents have the same symmetry properties.
To distinguish these contributions, one can use a subgap light, which induces only the rectification current.
Another way is to measure the spin current for samples with different carrier lifetimes (e.g. due to different impurity concentrations) and use the fact that only the injection current is proportional to the carrier lifetime.

%=========================================================
\section{Results}
\begin{figure}[!htb]
\centering
\includegraphics[width=1.0\columnwidth]{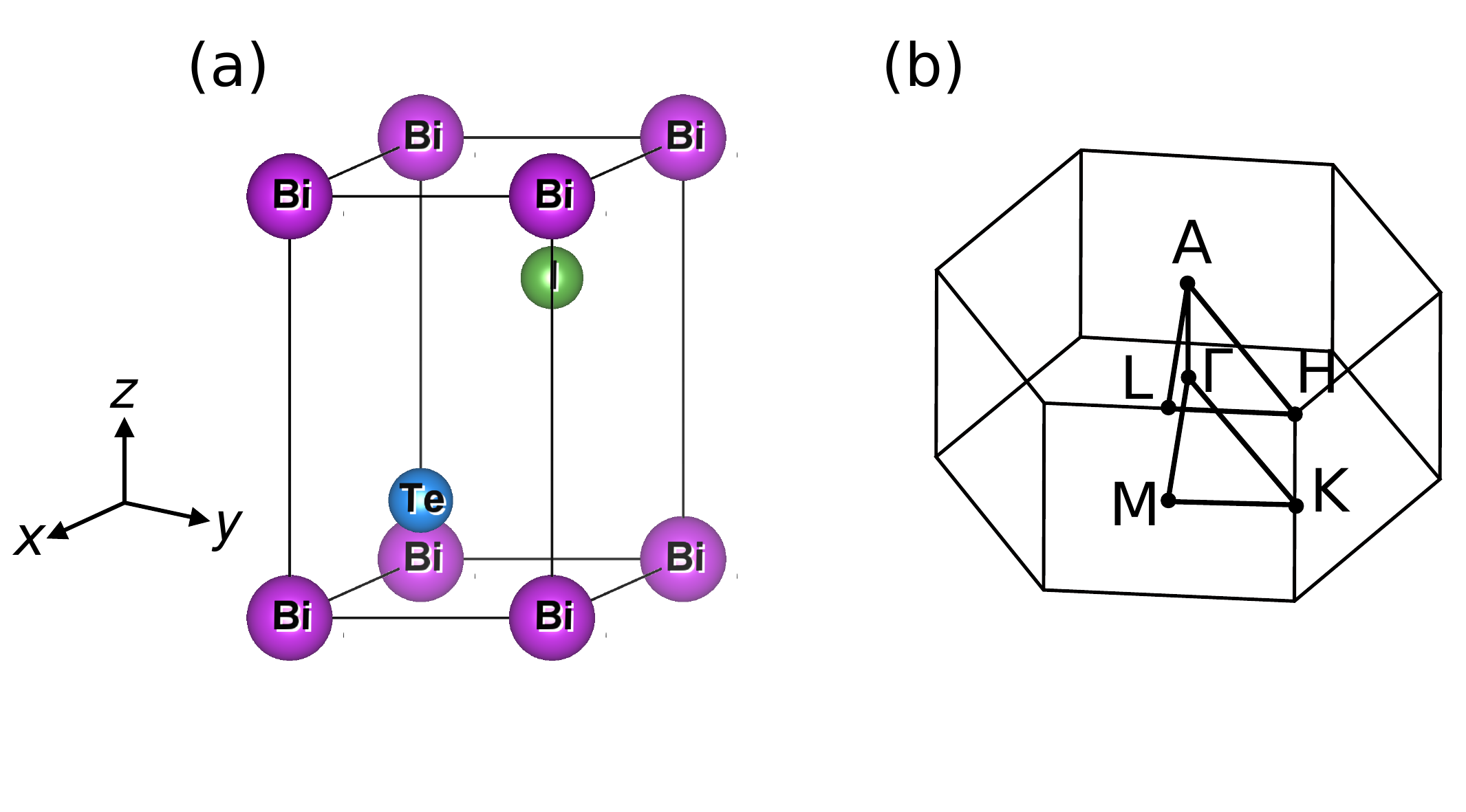}
\caption{
(a) Crystal structure of BiTeI.
(b) Brillouin zone of BiTeI.
The crystal structure is drawn using the software VESTA~\cite{2011MommaVESTA}.
}
\label{fig:structure}
\end{figure}

\begin{figure}[!htb]
\centering
\includegraphics[width=0.9\columnwidth]{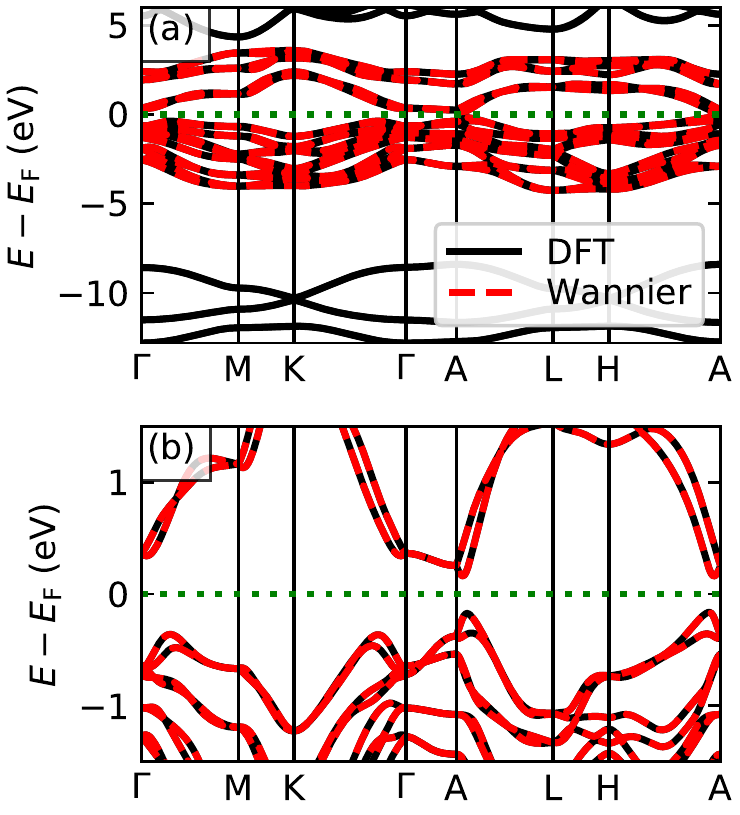}
\caption{
Electronic energy band structure of BiTeI calculated using density functional theory (DFT) and Wannier interpolation.
The green horizontal dotted lines indicate the center of the band gap.
}
\label{fig:band}
\end{figure}

Now, we apply our theory to the charge and spin photocurrent response of bulk BiTeI.
BiTeI is a non-magnetic polar direct-gap semiconductor with a giant Rashba-type spin splitting~\cite{2011IshizakaBiTeI}.
Here, we calculate the charge and spin photoconductivity of BiTeI using a Wannier-function-based {\it ab initio} tight-binding model~\cite{1997MarzariMLWF}.
Figure~\ref{fig:structure} shows the crystal structure and Brillouin zone of BiTeI, and Fig.~\ref{fig:band} shows the band structure of BiTeI.

\begin{figure*}[!htb]
\centering
\includegraphics[width=1.0\textwidth]{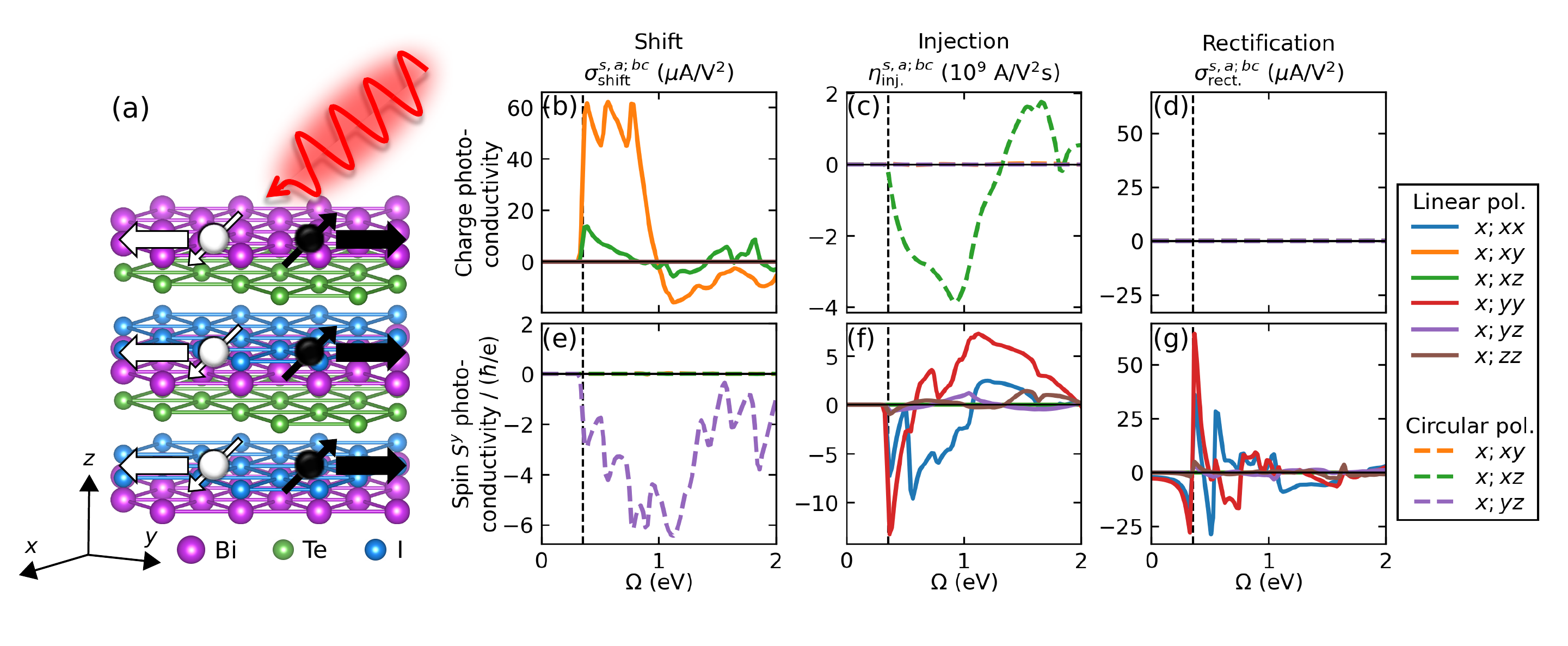}
\caption{
(a) Schematic illustration of light-induced spin currents in BiTeI.
(b-d) Charge and (e-g) spin $S^y$ photoconductivity of BiTeI.
The vertical dashed lines indicate the band gap of 0.35~eV.
The conductivity for linearly and circularly polarized light corresponds to the real symmetric and imaginary anti-symmetric components of the conductivity tensor, respectively.
}
\label{fig:sigma_all}
\end{figure*}

In Fig.~\ref{fig:sigma_all}, we show the charge and spin $S^y$ photoconductivity.
Spin $S^x$ and $S^z$ photoconductivities are shown in Fig.~\ref{fig:supp_sigma_xz}.
Since BiTeI is an insulator, the charge rectification current is zero in the entire frequency range.
In contrast, the spin rectification current is nonzero, even for subgap frequencies.
Therefore, our theory predicts pure spin currents in the subgap frequency regime.
We note that the spin rectification current diverges when $\Omega$ approaches the band gap in the limit of zero smearing.
In reality, this divergence will be regularized by the finite broadening of the bands arising from the finite lifetime of the electronic states.

The $\sigma^{y,x;yy}_\mathrm{rect.}$ component of the subgap spin photoconductivity, which describes the spin $S^y$ current along the $x$ direction with irradiation of light linearly polarized along the $y$ direction, has a magnitude around $20~\mathrm{\mu A/V^2} \hbar/e$ for an infrared light with frequency 0.30~eV.
The magnitude of this subgap spin response is larger than the calculated spin shift currents of collinear antiferromagnets BiFeO$_3$ and hematite~\cite{2013YoungSpin}.
This pure spin current could also be detected by using spin-to-charge conversion methods such as the inverse spin Hall effect~\cite{2006SaitohISHE}.

\begin{figure*}[htb]
\centering
\includegraphics[width=1.0\textwidth]{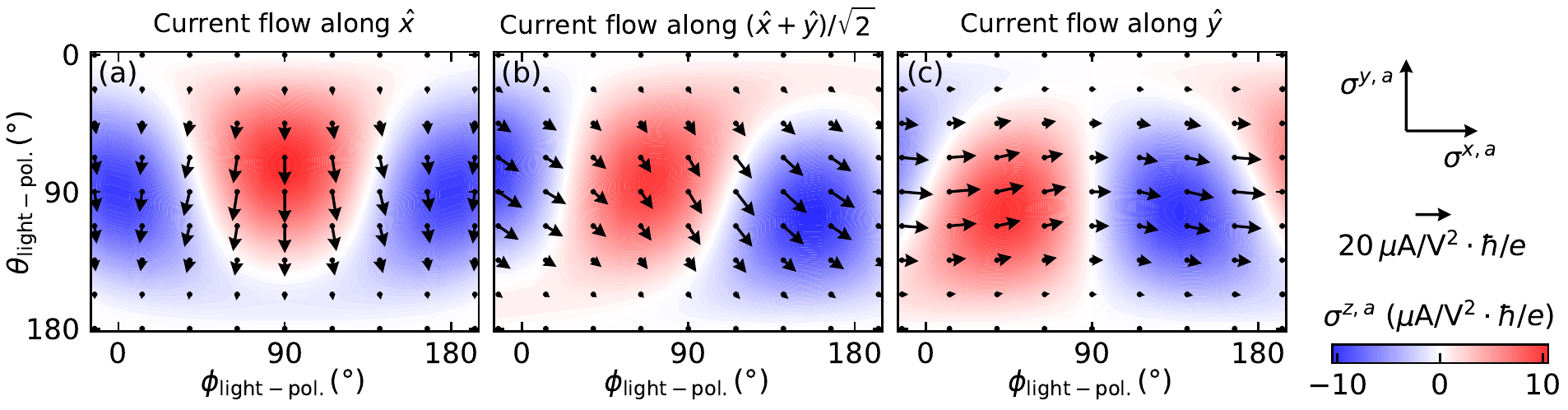}
\caption{
Spin polarization of the subgap spin rectification conductivity $\sigma^{s,a}$ [\myeqref{eq:polarization_sigma}] of BiTiI.
The arrows and colors indicate the in-plane and out-of-plane spin polarization of the current, respectively.
Light frequency is $\Omega=0.30$~eV, below the calculated band gap of 0.35~eV.
}
\label{fig:subgap_pol}
\end{figure*}

Remarkably, the polarization of the spin current can be tuned by the polarization of the light.
Figure~\ref{fig:subgap_pol} shows the spin polarization of the subgap spin conductivity as a function of light polarization angles:
\begin{equation} \label{eq:polarization_sigma}
    \sigma^{s,a}(\theta, \phi) = \sum_{b,\,c}\sigma^{s,a;bc}\, \hat{n}^b\, \hat{n}^c\,,
\end{equation}
where $\hat{n} = (\sin\theta \sin\phi,\, \sin\theta \cos\phi,\, \cos\theta)$ denotes the direction of the light polarization.
The energy of the incident photon is $\Omega = 0.30$~eV, below the calculated band gap of 0.35~eV.
Both the direction and magnitude of the spin polarization can be highly tuned by changing the light polarization.
Also, the spin polarization depends significantly on the direction of the spin current.
These remarkable tunabilities are not only scientifically important by themselves but may also open doors to novel spintronics applications.

%=========================================================
\section{Discussion}
For a material to produce a large spin photocurrent driven by spin-orbit coupling, it should consist of atoms with high atomic numbers, have a large structural asymmetry~\cite{2016TanReview}, and have a small band gap~\cite{2017CookShift}.
BiTeI, which is well known for the large Rashba effect and has a band gap of around 0.3~eV, is one of the materials satisfying all these criteria.
Monolayer SnS and SnSe, which contain the heavy Sn atoms and host large charge photocurrents~\cite{2017RangelBPVE}, are also good candidates.
One could search for other materials that can host large spin photocurrents based on these basic principles.

In this work, we considered the clean limit and found that only the spin injection current depends on the lifetime.
It has also been numerically tested that the relaxation time indeed has little effect on the spin and charge photocurrents~\cite{2019ZhangPhotocurrent,2020XuSpin}.
The effect of various scattering processes~\cite{1990Ivchenko,2002Ganichev,2006GanichevSpin,2020Budkin} beyond the simple phenomenological relaxation time approximation on spin photocurrents may be a subject for a future study.

Recently, Kaplan \textit{et al.}~\cite{2020KaplanSubgap} proposed that a subgap charge photocurrent exists in systems without time-reversal symmetry.
Within our theory, the subgap charge photocurrent is always zero, with or without the time-reversal symmetry.
We first note that since BiTeI has a time-reversal symmetry, the subgap charge photocurrent is zero in both theories.
The prediction of subgap photocurrents in a time-reversal symmetric system is unique to the spin-current response.
Also, in this work, we first take the clean limit of infinite quasiparticle lifetime and then take the DC limit $\omega \rightarrow 0$ as in the previous studies~\cite{2000Sipe,2019ParkerDiagram,2019ZhangPhotocurrent,2020DejuanPhotocurrent,2020GaoIntrinsicFS,2020AhnBPVE,2021WatanabePhotocurrent}, while Ref.~\cite{2020KaplanSubgap} takes the clean limit after the DC limit.
Thus, the two theories describe different physical situations.
Concretely, our results should be understood as describing the difference frequency generation scenario, where $\omega \gg 1/\tau$~\cite{2020DejuanPhotocurrent}.
Also, while the subgap charge photocurrent of Ref.~\cite{2020KaplanSubgap} depends on the detail of how the clean limit is taken, the subgap spin rectification photocurrent that we report is independent of such detail.
The relation between the subgap spin rectification current and the subgap charge photocurrent of Ref.~\cite{2020KaplanSubgap} is a subject of future study.

%=========================================================
\section{Conclusion}
In conclusion, we developed a complete theory of second-order spin photocurrents, which is generally applicable to systems with spin-orbit coupling or noncollinear magnetization.
The spin shift vector, which is a gauge-invariant quantity, governs the spin shift and rectification responses.
We found that subgap light can induce a DC spin-current response in a nonmagnetic insulator, which does not occur in the charge current case.
This phenomenon enables an optical generation of pure spin currents, whose spin polarization is highly tunable by the polarization of light or the flow direction of the spin current.
We applied our theory to BiTeI and found that the calculated subgap spin photoconductivity is large enough to be detectable in experiments.
Our study paves the way for theoretical and experimental studies of the nonlinear optical generation of spin currents.

\bigskip
\textit{Note added in proof} ---
Calculation of the spin shift~[\myeqref{eq:sigma_shift}] and rectification current~[\myeqref{eq:sigma_rec}], as well as the spin shift vector~[\myeqref{eq:spin_shift_vector}], requires a sum over an infinite number of bands.
In this work, we used the diagonal tight-binding approximation, which involves an artificial truncation of the bands.
Reference~\cite{2021LihmWFPT} presents an efficient method to calculate these quantities from first principles without any band-truncation error.

% =========================================================
\begin{acknowledgments}
This work was supported by the Creative-Pioneering Research Program through Seoul National University, Korean NRF No-2020R1A2C1014760, and the Institute for Basic Science (No. IBSR009-D1).
Computational resources have been provided by KISTI (KSC-2020-INO-0078).
\end{acknowledgments}

\appendix

% =========================================================
\section{Definition of the spin current} \label{sec:app_spin_current}
In presence of spin-orbit coupling, spin is not a conserved quantity.
Thus, the conventional definition of the spin-current operator~[\myeqref{eq:j_def}], which is used in most studies including ours, does not describe a conserved current.
Therefore, it is actually still debated whether this conventional, most widely used definition of the spin-current operator should be modified or not.

A modified definition of the spin-current operator as $j^{s,a}=d(r^a S^s)/dt$ has been proposed for an inversion-symmetric system in Ref.~\cite{2006ShiSpinCurrent}.
However, this definition gives a conserved current only if the “spin generation in the bulk is absent.”
In the case of light irradiation, which is a prerequisite for photocurrent response, the light-induced spin polarization is always present~\cite{2020XuSpin}
Moreover, the spin generation in the bulk through spin-orbit coupling is not forbidden in general when the bulk is inversion asymmetric, which is another prerequisite for having a second-order photocurrent response.
Thus, even this modified definition of spin current does not give a conserved current in the study of photocurrents.

Furthermore, there are some theoretical works claiming that the conventional definition should be used as is~\cite{2008SunSpinCurrent} because the non-conservation of spin current is a physical phenomenon with possible experimental outcomes.
Also, it has been estimated numerically that the difference between the conventional and modified spin currents will be on the order of 10 to 20\%, indicating that the computational results will not be qualitatively affected by the choice of the spin-current operator~\cite{2020XuSpin}.
Most importantly, the conventional definition has been tested against experiments in the context of spin Hall effects and gave good agreement on the spin Hall angle~\cite{2018QiaoSpinHall,2020WangSpinHall}.
Therefore, we used the conventional spin-current operator in our work.

We also note that to our knowledge all previous works on the spin photocurrents, including Refs.~\cite{2005BhatSpinInjection,2021FeiSpinInjection,2020XuSpin}, also used the conventional spin-current operator.

One interesting property of the conventional spin-current operator is that a nonzero spin current exists in equilibrium~\cite{2003Rashba}.
We note in passing that this equilibrium spin current is an intrinsic property of the spin current~\cite{2008SunSpinCurrent}.
As we study only the spin currents that occur in response to the external light, we did not consider the equilibrium spin current.

%========================================================
\section{Derivation of spin photoconductivity} \label{sec:app_derivation}
In this section, we derive the spin photocurrent conductivity.
We assume non-degenerate bands.
(For $\opmcP\opmcT$-symmetric cases, the recipe in Appendix B of Ref.~\cite{2021WatanabePhotocurrent} can be used.)

\subsection{Perturbative expansion of density matrix}
Under the independent particle approximation, the electron Hamiltonian reads
\begin{equation}
    \opH_0 = \sum_{\mb{k},m} \veps_\mk \opcd_\mk \opc_\mk.
\end{equation}
Here, $\veps_\mk$ is the band energy of state $m$ with crystal momentum $\mb{k}$, and $\opcd_\mk$ and $\opc_\mk$ are the electron creation and annihilation operators, respectively.
Using the Bloch theorem, one can write an eigenvalue equation for the band energy and the periodic part of the Bloch wavefunction:
\begin{equation}
    H_0(\mb{k}) \ket{u_\mk} = \veps_\mk \ket{u_\mk}.
\end{equation}
In the following, we omit the subscript $\mb{k}$ for brevity.

In the length gauge and the dipole approximation, the perturbation due to the electric field is given by
\begin{equation} \label{eq:supp_ham_dipole_def}
    \opV(t) = -q \hat{r}^a E^a(t),
\end{equation}
where $q$ is the charge of an electron, and $E^a$ and $r^a$ are the electric field and position operator along the direction $a$, respectively.
The sum over repeated superscripts is implied.
Here, $\hat{r}^a$ is the position operator, whose matrix element is~\cite{1962Blount}
\begin{equation} \label{eq:supp_dipole_def}
    r^a_{mn} = i \delta_{mn} \partial^a + \xi^a_{mn}
\end{equation}
where we defined $\partial^a = \partial / \partial k_a$ and the Berry connection $\xi^a_{mn} = i \braket{u_m}{\partial^a u_n}$.

The Schr\"odinger equation for the one-particle reduced density matrix $\rho_{mn}$ is~\cite{2017VenturaGauge}
\begin{equation}
    i \hbar \frac{d \rho_{mn}(t)}{dt}
    = \hbar \omega_{mn} \rho_{mn}(t)
    - q E^a(t) [r^a, \rho(t)]_{mn}.
\end{equation}
Defining the Fourier transformation as
\begin{equation}
    \rho_{ab}(t) = \int \frac{d\omega}{2\pi} e^{-i\omega t} \rho_{ab}(\omega),
\end{equation}
the Schr\"odinger equation in the frequency domain becomes
\begin{equation}
    \hbar( \omega - \omega_{mn}) \rho_{mn}(\omega)
    = -q \int \frac{d\Omega}{2\pi} E^a(\Omega) \comm{r^a}{\rho(\omega - \Omega)}_{mn}.
\end{equation}

Regarding the electric field as a perturbation, one can expand $\rho$ as
\begin{equation}
    \rho(\omega) = \sum_n \rho^{(n)}(\omega),
\end{equation}
where $\rho^{(n)}$ is of order $\mathcal{O}(E^n).$
The zeroth-order term is the Fermi-Dirac distribution:
\begin{equation}
    \rho^{(0)}_{ab}(\omega) = 2\pi \delta(\omega) f_a \delta_{ab}.
\end{equation}
Using the perturbative expansion of the reduced density matrix, the $n$-th order charge and spin current responses can be calculated as
\begin{align} \label{eq:supp_Jn_rhon}
    J^{s,a}_{(n)}(\omega)
    = \frac{1}{V} \sum_{\mb{k},m,n} q j^{s,a}_{mn} \rho^{(n)}_{nm}(\omega).
\end{align}
Here, $V$ is the volume of the system.

In this work, we focus on the second-order response, which is the lowest order where a DC current response can occur under AC driving fields.
We define the second-order conductivity tensor $\sigma^{s,a;bc}\omegaall$ as
\begin{align} \label{eq:supp_sigma2_def}
    J^{s,a}_{(2)}(\omega)
    = \int \frac{d\omega_1 d\omega_2}{(2\pi)^2} &\sigma^\sabc\omegaall
    E^b(\omega_1) E^c(\omega_2) \nnnl
    &\times 2\pi \delta(\omega - \omega_1 - \omega_2).
\end{align}
To investigate the DC photocurrent, we set
\begin{equation} \label{eq:supp_dc_limit}
    \omega_1 = -\Omega, \quad \omega_2 = \Omega + \omega
\end{equation}
and take the DC limit $\omega \rightarrow 0$.
In other words, we study $\sigma^{s,a;bc}(0;-\Omega,\Omega)$.
We assume a clean system with an infinite carrier lifetime as done in previous studies on charge-current responses~\cite{2000Sipe,2019ParkerDiagram,2019ZhangPhotocurrent,2020DejuanPhotocurrent,2020GaoIntrinsicFS,2020AhnBPVE,2021WatanabePhotocurrent}.

Watanabe and Yanase~\cite{2021WatanabePhotocurrent} showed that the matrix elements of the second-order reduced density operator can be divided into terms originating from the intraband (i) and interband (e) position operators as
\begin{equation} \label{eq:supp_rho_contributions}
    \rho_{mn}^{(2)}
    = \rho_{mn}^{\rm (ii)}
    + \rho_{mn}^{\rm (ei)}
    + \rho_{mn}^{\rm (ie)}
    + \rho_{mn}^{\rm (ee)}.
\end{equation}
Following Ref.~\cite{2021WatanabePhotocurrent}, we define the contribution of each term of \myeqref{eq:supp_rho_contributions} to the conductivity as $\sigma^\sabc_{(X)}$ for $X=\mathrm{ii}$, ei, ie, and ee, and we calculate each term.
The physical phenomena are quite different from the case of charge currents if we apply this reduced density matrix to the calculation of spin currents.

\begin{widetext}
Let us define $\rho_{mn}^{bc\,(X)}\omegaall$ to satisfy
\begin{equation} \label{eq:supp_rho_Efield}
    \rho_{mn}^{(X)} = \frac{1}{2} \int \frac{d\omega_1 d\omega_2}{(2\pi)^2} E^b(\omega_1) E^c(\omega_2) \rho_{mn}^{bc\,(X)}\omegaall 2\pi\delta(\omega - \omega_1 - \omega_2) + [(b,\omega_1) \leftrightarrow (c,\omega_2)]
\end{equation}
for $X=2$, ii, ei, ie, and ee.
Then, $\rho_{mn}^{bc\,(2)}\omegaall$ can be decomposed as follows~\cite{2021WatanabePhotocurrent}.
\begin{equation} \label{eq:supp_rho_bc_contributions}
    \rho_{mn}^{bc\,(2)}\omegaall
    = \sum_{X = \mathrm{ii,\,ei,\,ie,\,ee}} \rho_{mn}^{bc\,(X)}\omegaall
\end{equation}
\begin{equation} \label{eq:supp_rho_ii_def}
    \rho_{mn}^{bc\,\mathrm{(ii)}}\omegaall
    = (-iq)^2 \delta_{mn} d^\omega d^{\omega - \omega_1} \der{b}\der{c}f_m
\end{equation}
\begin{equation} \label{eq:supp_rho_ei_def}
    \rho_{mn}^{bc\,\mathrm{(ei)}}\omegaall
    = -iq^2 d^\omega_{mn} d^{\omega - \omega_1} \xi^b_{mn} \der{c} f_{mn}
\end{equation}
\begin{equation} \label{eq:supp_rho_ie_def}
    \rho_{mn}^{bc\,\mathrm{(ie)}}\omegaall
    = -iq^2 d^\omega_{mn}
    [ \der{b} (d^{\omega - \omega_1}_{mn} f_{mn} \xi^c_{mn}) - i(\xi^b_{mm} - \xi^b_{nn}) d^{\omega - \omega_1}_{mn} f_{mn} \xi^c_{mn}]
\end{equation}
\begin{equation} \label{eq:supp_rho_ee_def}
    \rho_{mn}^{bc\,\mathrm{(ee)}}\omegaall
    = q^2 \sum_{\substack{p\neq m,n}} d^\omega_{mn}
    [ d^{\omega - \omega_1}_{pn} \xi^b_{mp} \xi^c_{pn} f_{np}
    - d^{\omega - \omega_1}_{mp} \xi^b_{pn} \xi^c_{mp} f_{pm}]
\end{equation}
Here, we defined
\begin{equation}
    d^\omega_{mn} = \frac{1}{\hbar \omega + i0^+ - \hbar \omega_{mn}}
\end{equation}
with $0^+$ a positive infinitesimal value that ensures causality of the response.
We write the diagonal components $d^\omega_{mm}$ as $d^\omega$ since its value does not depend on $m$.
Note that the constraint $p\neq m,n$ is included in the definition of $\rho_{mn}^{bc\,\mathrm{(ee)}}$ [\myeqref{eq:supp_rho_ee_def}] to make sure that the intraband effect is excluded.
The intraband effect is considered in $\rho_{mn}^{bc\,\mathrm{(ie)}}$ [\myeqref{eq:supp_rho_ie_def}].
Equations (\ref{eq:supp_rho_ii_def}-\ref{eq:supp_rho_ee_def}) corresponds to Eqs.~(16-19) of Ref.~\cite{2021WatanabePhotocurrent}.

Using \myeqref{eq:supp_Jn_rhon}, \myeqref{eq:supp_sigma2_def}, and \myeqref{eq:supp_rho_Efield} we find the following expression for the second-order spin photoconductivity.
\begin{equation}
    \sigma^\sabc_{(X)}\omegaall = \frac{q}{2V} \sum_{\mb{k},m,n} j_{mn}^{s,a} \rho_{nm}^{bc\,(X)}\omegaall + [(b,\omega_1) \leftrightarrow (c,\omega_2)].
\end{equation}
Using the expressions for the second-order density matrix elements [Eqs.~(\ref{eq:supp_rho_ii_def}-\ref{eq:supp_rho_ee_def})], the second-order spin photoconductivity can be computed as follows.
\begin{equation} \label{eq:supp_sigma_contributions}
    \sigma^\sabc\omegaall
    = \sigma^\sabc_\subii\omegaall
    + \sigma^\sabc_\subei\omegaall
    + \sigma^\sabc_\subie\omegaall
    + \sigma^\sabc_\subee\omegaall
\end{equation}
\begin{equation} \label{eq:supp_sigma_ii_def}
    \sigma^\sabc_\subii\omegaall
    = \frac{q^3}{2V} \sum_{\mb{k},m} -j^{s,a}_{mm} d^\omega d^{\omega_2} \der{b} \der{c} f_m
    + [(b,\omega_1) \leftrightarrow (c,\omega_2)]
\end{equation}
\begin{equation} \label{eq:supp_sigma_ei_def}
    \sigma^\sabc_\subei\omegaall
    = \frac{q^3}{2V} \sum_{\mb{k},m,n} -i j^{s,a}_{mn} d^\omega_{nm} d^{\omega_2} \xi^b_{nm} \der{c} f_{nm}
    + [(b,\omega_1) \leftrightarrow (c,\omega_2)]
\end{equation}
\begin{equation} \label{eq:supp_sigma_ie_def}
    \sigma^\sabc_\subie\omegaall
    = \frac{q^3}{2V} \sum_{\mb{k},m,n} -i j^{s,a}_{mn} d^\omega_{nm}
    [ \der{b} (d^{\omega_2}_{nm} f_{nm} \xi^c_{nm}) - i(\xi^b_{nn} - \xi^b_{mm}) d^{\omega_2}_{nm} f_{nm} \xi^c_{nm}]
    + [(b,\omega_1) \leftrightarrow (c,\omega_2)]
\end{equation}
\begin{equation} \label{eq:supp_sigma_ee_def}
    \sigma^\sabc_\subee\omegaall
    = \frac{q^3}{2V} \sum_{\substack{\mb{k},m,n,p \\ p\neq m,n}} j^{s,a}_{mn} d^\omega_{nm}
    [ d^{\omega_2}_{pm} \xi^b_{np} \xi^c_{pm} f_{mp}
    - d^{\omega_2}_{np} \xi^b_{pm} \xi^c_{np} f_{pn}]
    + [(b,\omega_1) \leftrightarrow (c,\omega_2)]
\end{equation}

Now, we take the DC limit~[\myeqref{eq:supp_dc_limit}] and analyze each term.

\subsection{Drude current}
First, we show that the DC limit of $\sigma^\sabc_\subii\omegaall$ is the Drude conductivity~[\myeqref{eq:sigma_drude}].
\begin{align} \label{eq:supp_sigma_drude_derive}
    \sigma^\sabc_\subii(0;-\Omega,\Omega)
    =& \lim_{\omega \rightarrow 0} \frac{q^3}{2\hbar^2 V\omega} \sum_{\mb{k},m} -j^{s,a}_{mm} \left(\frac{1}{\Omega + \omega} + \frac{1}{-\Omega} \right) \der{b} \der{c} f_m \nnnl
    =& \lim_{\omega \rightarrow 0} \frac{q^3}{2\hbar^2 V\omega} \sum_{\mb{k},m} j^{s,a}_{mm} \frac{\omega}{(\Omega + \omega)\Omega} \der{b} \der{c} f_m \nnnl
    =& \frac{q^3}{2\hbar^2 V\Omega^2} \sum_{\mb{k},m} j^{s,a}_{mm} \der{b} \der{c} f_m \nnnl
    =& \sigma^\sabc_\mathrm{Drude}(\Omega)
\end{align}

\subsection{Berry curvature dipole current}
Next, we show that the DC limit of $\sigma^\sabc_\subei\omegaall$ is the Berry curvature dipole current contribution~[\myeqref{eq:sigma_BCD}].
\begin{align} \label{eq:supp_sigma_bcd_derive}
    \sigma^\sabc_\subei(0;-\Omega,\Omega)
    =& \frac{q^3}{2 \hbar^2 V\Omega} \sum_{\substack{\mb{k},m,n\\m \neq n}} \frac{j^{s,a}_{mn} v^b_{nm}}{\omega_{mn}^2} \der{c} f_{nm}
    + [(b,-\Omega) \leftrightarrow (c,\Omega)] \nnnl
    =& \frac{q^3}{2 \hbar^2 V\Omega} \sum_{\substack{\mb{k},m,n\\m \neq n}} \left( \frac{j^{s,a}_{nm} v^b_{mn}}{\omega_{mn}^2} \der{c} f_{m}
    - \frac{j^{s,a}_{mn} v^b_{nm}}{\omega_{mn}^2} \der{c} f_{m} \right)
    - (b \leftrightarrow c) \nnnl
    =& \frac{-iq^3}{\hbar^2 V\Omega} \sum_{\substack{\mb{k},m,n\\m \neq n}} \frac{\Im(j^{s,a}_{mn} v^b_{nm})}{\omega_{mn}^2} \der{c} f_{m}
    - (b \leftrightarrow c) \nnnl
    =& \sigma^\sabc_\mathrm{BCD}(\Omega)
\end{align}

Defining the spin Berry curvature~\cite{2008GuoSHE}
\begin{equation}
    \Omega^{s,ab}_{m} = -2 \Im \sum_{n \neq m} \frac{j^{s,a}_{mn} v^b_{nm}}{\omega_{mn}^2},
\end{equation}
one can rewrite \myeqref{eq:supp_sigma_bcd_derive} as
\begin{align} \label{eq:supp_sigma_BCD_Berry}
    \sigma^\sabc_\mathrm{BCD}
    =& \frac{i q^3}{2\hbar^2 V\Omega} \sum_{\mb{k},m} \Omega^{s,ab}_{m} \der{c} f_{m}
    - (b \leftrightarrow c)
    = \frac{-i q^3}{2\hbar^2 V\Omega} \sum_{\mb{k},m} f_{m}
    (\der{c} \Omega^{s,ab}_{m} - \der{b} \Omega^{s,ac}_{m}).
\end{align}
Equation~\eqref{eq:supp_sigma_BCD_Berry} clearly reveals that $\sigma^\sabc_\mathrm{BCD}$ is proportional to the momentum space dipole of the spin Berry curvature.

\subsection{Injection, shift, rectification currents}
Finally, we derive the formulas for the injection, shift, and rectification photoconductivities [Eqs.~(\ref{eq:sigma_inj}-\ref{eq:sigma_rec})] by showing that the sum of $\sigma^\sabc_\subie$ and $\sigma^\sabc_\subee$ can be reorganized as the sum of injection, shift, and rectification currents.

First, let us consider $\sigma^\sabc_\mathrm{(ee,d)}$, which is the component of $\sigma^\sabc_\subee$ [\myeqref{eq:supp_sigma_ee_def}] with $m=n$:
\begin{align}
    \sigma^\sabc_\mathrm{(ee,d)}\omegaall
    =& \frac{q^3}{2V} \sum_{\mb{k},m,p} j^{s,a}_{mm} d^\omega
    [ d^{\omega_2}_{pm} \xi^b_{mp} \xi^c_{pm} f_{mp}
    - d^{\omega_2}_{mp} \xi^b_{pm} \xi^c_{mp} f_{pm}]
    + [(b,\omega_1) \leftrightarrow (c,\omega_2)] \nnnl
    =& \frac{q^3}{2 \hbar V\omega} \sum_{\mb{k},m,p}
    \Delta j^{s,a}_{mp} d^{\omega_2}_{pm} \xi^b_{mp} \xi^c_{pm} f_{mp}
    + [(b,\omega_1) \leftrightarrow (c,\omega_2)] \nnnl
    =& \frac{q^3}{2 \hbar V\omega} \sum_{\mb{k},m,p}
    \Delta j^{s,a}_{mp} \xi^b_{mp} \xi^c_{pm} f_{mp} (d^{\omega_1}_{mp} + d^{\omega_2}_{pm}).
\end{align}
Since we are interested in $\sigma^{s,a;bc}(0;-\Omega,\Omega)$, using
\begin{align}
    d^{\omega_1}_{mp} + d^{\omega_2}_{pm}
    =& \frac{1}{-\hbar \Omega + i0^+ - \hbar \omega_{mp}}
    + \frac{1}{\hbar (\Omega + \omega) + i0^+ - \hbar \omega_{pm}} \nnnl
    =& -2\pi i \delta(\hbar \Omega + \hbar \omega_{mp})
    - \frac{\hbar \omega}{(\hbar \Omega + \hbar \omega_{mp})^2}
    + \mathcal{O}(\omega^2),
\end{align}
we find
\begin{align} \label{eq:supp_sigma_eed_final}
    \sigma^\sabc_\mathrm{(ee,d)}(\omega)
    =& \frac{-i \pi q^3}{\hbar^2 V\omega} \sum_{\mb{k},m,n}
    \Delta j^{s,a}_{mn} \xi^b_{mn} \xi^c_{nm} f_{mn}
    \delta(\Omega + \omega_{mn})
    - \frac{q^3}{2V} \sum_{\mb{k},m,n}
    \Delta j^{s,a}_{mn} \xi^b_{mn} \xi^c_{nm} f_{mn}
    \frac{1}{(\hbar \Omega + \hbar\omega_{mn})^2}
    + \mathcal{O}(\omega) \nnnl
    =& \frac{-i \pi q^3}{\hbar^2 V\omega \Omega^2} \sum_{\mb{k},m,n} f_{mn}
    \Delta j^{s,a}_{mn} v^b_{mn} v^c_{nm} \delta(\Omega + \omega_{mn})
    - \frac{q^3}{2\hbar^2V} \sum_{\substack{\mb{k},m,n\\m\neq n}} \frac{f_{mn}}{\omega_{mn}^2}
    \frac{\Delta j^{s,a}_{mn} v^b_{mn} v^c_{nm}}{(\Omega + \omega_{mn})^2}
    + \mathcal{O}(\omega).
\end{align}

The first term of \myeqref{eq:supp_sigma_eed_final}, which contains the $1/\omega$ divergence, is the injection-current contribution:
\begin{equation} \label{eq:supp_sigma_inj_omega}
    \sigma^\sabc_\mathrm{inj.}(\omega) = \frac{-i \pi q^3}{\hbar^2 V\omega \Omega^2} \sum_{\mb{k},m,n} f_{mn} \Delta j^{s,a}_{mn} v^b_{mn} v^c_{nm}
    \delta(\Omega + \omega_{mn}).
\end{equation}
The $1/\omega$ divergence in the injection current [\myeqref{eq:supp_sigma_inj_omega}] indicates that the photocurrent increases linearly with time.
This divergence can be avoided by introducing a phenomenological relaxation time $\tau$~\cite{2018PassosVelocitygauge}, which is justified by calculations based on the Floquet formalism~\cite{2017deJuanCPVE}.
The DC injection conductivity then becomes
\begin{align} \label{eq:supp_sigma_inj_tau}
    &\sigma^\sabc_\mathrm{inj.,\tau}
    = \tau \eta^\sabc_\mathrm{inj.}
    = -\tau\frac{\pi q^3}{\hbar^2 \Omega^2 V} \sum_{\mb{k},m,n} f_{mn} \Delta j^{s,a}_{mn} v^b_{mn} v^c_{nm} \delta(\Omega + \omega_{mn}).
\end{align}

The second term of \myeqref{eq:supp_sigma_eed_final} is a part of the rectification current:
\begin{equation} \label{eq:supp_sigma_rect_1}
    \sigma^\sabc_\mathrm{rect.,1} = - \frac{q^3}{2 \hbar^2 V} \sum_{\substack{\mb{k},m,n\\m\neq n}} \frac{f_{mn}}{\omega_{mn}^2}
    \frac{\Delta j^{s,a}_{mn} v^b_{mn} v^c_{nm}}{(\Omega + \omega_{mn})^2}
\end{equation}

Next, let us consider $\sigma^\sabc_\subie$.
In the DC limit, one finds
\begin{align} \label{eq:supp_sigma_ie_1}
    \sigma^\sabc_\subie
    =& \frac{q^3}{2V} \sum_{\substack{\mb{k},m,n\\m\neq n}} -i j^{s,a}_{mn} \frac{1}{\veps_{mn}}
    [ \partial^b (d^{\Omega}_{nm} f_{nm} \xi^c_{nm}) - i(\xi^b_{nn} - \xi^b_{mm}) d^{\Omega}_{nm} f_{nm} \xi^c_{nm}]
    + [(b,-\Omega) \leftrightarrow (c, \Omega)] \nnnl
    =& \frac{q^3}{2V} \sum_{\substack{\mb{k},m,n\\m\neq n}} i \left[
    (\partial^b j^{s,a}_{mn})
    - j^{s,a}_{mn}\frac{\Delta^b_{mn}}{\omega_{mn}}
    + i j^{s,a}_{mn} (\xi^b_{nn} - \xi^b_{mm})
    \right]
    \frac{1}{\hbar \omega_{mn}} d^{\Omega}_{nm} f_{nm} \xi^c_{nm}
    + [(b,-\Omega) \leftrightarrow (c, \Omega)] \nnnl
    =& \frac{i q^3}{2\hbar V} \sum_{\substack{\mb{k},m,n\\m\neq n}} \left( j^{s,a;b}_{mn}
    - \frac{j^{s,a}_{mn} \Delta^b_{mn}}{\omega_{mn}} \right)
    \frac{f_{nm}}{\omega_{mn}} d^{\Omega}_{nm} \xi^c_{nm}
    + [(b,-\Omega) \leftrightarrow (c, \Omega)].
\end{align}
In the second equality, we used partial integration.
We also defined
\begin{equation}
    \Delta^b_{mn} \equiv v^b_{mm} - v^b_{nn}
\end{equation}
and the generalized derivative
\begin{equation}
    j^{s,a;b}_{mn} \equiv \partial^b j^{s,a}_{mn} - i j^{s,a}_{mn} (\xi^b_{mm} - \xi^b_{nn}).
\end{equation}
Note that the generalized derivative $j^{s,a;b}_{mn}$ is an actual derivative with respect to the crystal momentum, while the spin-velocity derivative $d^{s,b;a}_{mn}$ is not.

Last, let us consider the remaining component of $\sigma^\sabc_\subee$ with $m \neq n$, which we denote as $\sigma^\sabc_\mathrm{(ee,od)}$.
In the DC limit, $\sigma^\sabc_\mathrm{(ee,od)}$ becomes
\begin{align} \label{eq:supp_sigma_eeod_1}
    \sigma^\sabc_\mathrm{(ee,od)}
    =& \frac{q^3}{2V} \sum_{\substack{\mb{k},m,n,p \\ m\neq n, p\neq m,n}} j^{s,a}_{mn} \frac{1}{\hbar \omega_{mn}}
    ( d^{\Omega}_{pm} \xi^b_{np} \xi^c_{pm} f_{mp}
    - d^{\Omega}_{np} \xi^b_{pm} \xi^c_{np} f_{pn})
    + [(b,-\Omega) \leftrightarrow (c,\Omega)] \nnnl
    =& \frac{q^3}{2\hbar V} \sum_{\substack{\mb{k},m,n}}
    \sum_{\substack{p \\ p\neq m,n}} \left( \frac{j^{s,a}_{mp} \xi^b_{pn}}{\omega_{mp}}
    + \frac{j^{s,a}_{pn} \xi^b_{mp}}{\omega_{np}} \right)
    f_{mn} d^{\Omega}_{nm} \xi^c_{nm}
    + [(b,-\Omega) \leftrightarrow (c,\Omega)].
\end{align}
In the second equality, we changed the dummy indices $(m,n,p)$ to $(m,p,n)$ and $(p,n,m)$ for the first and second term in the parentheses, respectively.

To add \myeqref{eq:supp_sigma_ie_1} and \myeqref{eq:supp_sigma_eeod_1}, we use the following formula:
\begin{align} \label{eq:supp_sigma_jab_equality}
    &\frac{j^{s,a;b}_{mn}}{\omega_{mn}}
    - j^{s,a}_{mn}\frac{\Delta^b_{mn}}{\omega_{mn}^2}
    +i \sum_{\substack{p \\ p\neq m,n}} \left( \frac{j^{s,a}_{mp} \xi^b_{pn}}{\omega_{mp}}
    + \frac{j^{s,a}_{pn} \xi^b_{mp}}{\omega_{np}} \right) \nnnl
    =& \frac{j^{s,ab}_{mn}}{\omega_{mn}}
    -\frac{i}{\omega_{mn}} \sum_{p\neq n} j^{s,a}_{mp} \xi^b_{pn}
    +\frac{i}{\omega_{mn}} \sum_{p\neq m} \xi^b_{mp} j^{s,a}_{pn} 
    - j^{s,a}_{mn}\frac{\Delta^b_{mn}}{\omega_{mn}^2}
    +i \sum_{p\neq m,n} \frac{j^{s,a}_{mp} \xi^b_{pn}}{\omega_{mp}}
    +i \sum_{p\neq m,n} \frac{j^{s,a}_{pn} \xi^b_{mp}}{\omega_{np}} \nnnl
    =& \frac{j^{s,ab}_{mn}}{\omega_{mn}}
    -i \frac{\Delta j^{s,a}_{mn} \xi^b_{mn}}{\omega_{mn}}
    -\frac{j^{s,a}_{mn} \Delta^b_{mn}}{\omega_{mn}^2}
    +\frac{i}{\omega_{mn}} \sum_{p\neq m,n} \frac{j^{s,a}_{mp} \xi^b_{pn} \omega_{pn}}{\omega_{mp}}
    +\frac{i}{\omega_{mn}} \sum_{p\neq m,n} \frac{j^{s,a}_{pn} \xi^b_{mp} \omega_{mp}}{\omega_{np}} \nnnl
    =& \frac{j^{s,ab}_{mn}}{\omega_{mn}}
    - \frac{\Delta j^{s,a}_{mn} v^b_{mn}}{\omega_{mn}^2}
    - \frac{j^{s,a}_{mn} \Delta^b_{mn}}{\omega_{mn}^2}
    + \frac{1}{\omega_{mn}} \sum_{p\neq m,n} \frac{j^{s,a}_{mp} v^b_{pn}}{\omega_{mp}}
    + \frac{1}{\omega_{mn}} \sum_{p\neq m,n} \frac{v^b_{mp} j^{s,a}_{pn}}{\omega_{np}} \nnnl
    =& \frac{1}{\omega_{mn}} \left( j^{s,ab}_{mn}
    + \sum_{p\neq m} \frac{j^{s,a}_{mp} v^b_{pn}}{\omega_{mp}}
    + \sum_{p\neq n} \frac{v^b_{mp} j^{s,a}_{pn}}{\omega_{np}} \right)
    - \frac{\Delta j^{s,a}_{mn} v^b_{mn}}{\omega_{mn}^2} \nnnl
    =& \frac{d^{s,b;a}_{mn}}{\omega_{mn}}
    - \frac{\Delta j^{s,a}_{mn} v^b_{mn}}{\omega_{mn}^2}.
\end{align}
From Eqs.~(\ref{eq:supp_sigma_ie_1}, \ref{eq:supp_sigma_eeod_1}, \ref{eq:supp_sigma_jab_equality}), we find
\begin{align} \label{eq:supp_sigma_eeod_2}
    &\sigma^\sabc_\subie + \sigma^\sabc_\mathrm{(ee,od)} \nnnl
    =& \frac{i q^3}{2 \hbar V} \sum_{\substack{\mb{k},m,n\\m\neq n}} \left[ \frac{j^{s,a;b}_{mn}}{\omega_{mn}}
    - j^{s,a}_{mn}\frac{\Delta^b_{mn}}{\omega_{mn}^2}
    + i\sum_{\substack{c \\ c\neq a,b}} \left( \frac{j^{s,a}_{mp} \xi^b_{pn}}{\omega_{mp}}
    + \frac{j^{s,a}_{pn} \xi^b_{mp}}{\omega_{np}} \right) \right]
    f_{nm} d^{\Omega}_{nm} \xi^c_{nm}
    + [(b,-\Omega) \leftrightarrow (c,\Omega)] \nnnl
    =& \frac{i q^3}{2 \hbar V} \sum_{\substack{\mb{k},m,n\\m\neq n}} \left( \frac{d^{s,b;a}_{mn}}{\omega_{mn}}
    - \frac{\Delta j^{s,a}_{mn} v^b_{mn}}{\omega_{mn}^2} \right)
    f_{nm} d^{\Omega}_{nm} \xi^c_{nm}
    + [(b,-\Omega) \leftrightarrow (c,\Omega)].
\end{align}

Now, let us separate $d^{\Omega}_{nm}$ into the delta function part and the principal value part:
\begin{equation}
    d^{\Omega}_{nm}
    = \frac{1}{\hbar\Omega + i0^+ + \hbar \omega_{mn}}
    = -i\pi\delta(\hbar \Omega + \hbar \omega_{mn})
    + \mathrm{P} \frac{1}{\hbar(\Omega + \omega_{mn})}.
\end{equation}
The delta function part is the shift current:
\begin{align} \label{eq:supp_shift}
    \sigma^\sabc_\mathrm{shift}
    =& \frac{i \pi q^3}{2 \hbar^2 V} \sum_{\substack{\mb{k},m,n\\m\neq n}} \left(
    \frac{d^{s,b;a}_{mn} v^c_{nm}}{\omega_{mn}^2}
    - \frac{\Delta j^{s,a}_{mn} v^b_{mn} v^c_{nm}}{\omega_{mn}^3} \right)
    f_{nm} \delta(\Omega + \omega_{mn}) 
    + [(b,-\Omega) \leftrightarrow (c,\Omega)] \nnnl
    =& \frac{i \pi q^3}{2\hbar^2 V \Omega^2} \sum_{\mb{k},m,n}
    (d^{s,b;a}_{mn} v^c_{nm} - d^{s,c;a}_{nm} v^b_{mn})
    f_{nm} \delta(\Omega + \omega_{mn}).
\end{align}
The principal value part is
\begin{align} \label{eq:supp_sigma_rect_2}
    \sigma^\sabc_\mathrm{rect.,2}
    =& \frac{q^3}{2 \hbar^2 V} \sum_{\substack{\mb{k},m,n\\m\neq n}} f_{mn} \left( \frac{d^{s,b;a}_{mn} v^c_{nm}}{\omega_{mn}^2}
    - \frac{\Delta j^{s,a}_{mn} v^b_{mn} v^c_{nm}}{\omega_{mn}^3} \right)
    \mathrm{P} \frac{1}{\Omega + \omega_{mn}}
    + [(b,-\Omega) \leftrightarrow (c,\Omega)] \nnnl
    =& \frac{q^3}{2 \hbar^2 V} \sum_{\substack{\mb{k},m,n\\m\neq n}} f_{mn} \left( 
    \frac{d^{s,b;a}_{mn} v^c_{nm} + d^{s,c;a}_{nm} v^b_{mn}}{\omega_{mn}^2}
    - \frac{2\Delta j^{s,a}_{mn} v^b_{mn} v^c_{nm}}{\omega_{mn}^3} 
    \right)
    \mathrm{P} \frac{1}{\Omega + \omega_{mn}}
\end{align}

Adding \myeqref{eq:supp_sigma_rect_1} and \myeqref{eq:supp_sigma_rect_2}, we obtain the rectification current:
\begin{align} \label{eq:supp_sigma_rect}
    \sigma^\sabc_\mathrm{rect.}
    =& \sigma^\sabc_\mathrm{rect., 1} + \sigma^\sabc_\mathrm{rect., 2} \nnnl
    =& \frac{q^3}{2 \hbar^2 V} \sum_{\substack{\mb{k},m,n\\m\neq n}} f_{mn}
    \left[ \left( 
    d^{s,b;a}_{mn} v^c_{nm} + d^{s,c;a}_{nm} v^b_{mn}
    - \frac{2\Delta j^{s,a}_{mn} v^b_{mn} v^c_{nm}}{\omega_{mn}} 
    \right) \frac{1}{\omega_{mn}^2} \mathrm{P} \frac{1}{\Omega + \omega_{mn}}
    - \frac{\Delta j^{s,a}_{mn} v^b_{mn} v^c_{nm}}{\omega_{mn}^2(\Omega + \omega_{mn})^2} \right].
\end{align}
\end{widetext}

%========================================================
\section{Symmetry analysis of photoconductivity} \label{sec:supp_symmetry}

In this section, we study the symmetry properties of the second-order photoconductivity tensors.
The real symmetric (imaginary anti-symmetric) component of the photoconductivity tensor, $\Re(\sigma^\sabc + \sigma^{s,a;cb})$ ($\Im(\sigma^\sabc - \sigma^{s,a;cb})$) corresponds to the generation of a real-valued current under linearly (circularly) polarized light.
In other words, the photoconductivity tensor for linearly (circularly) polarized light is symmetric (anti-symmetric) with respect to the exchange of indices $b$ and $c$.
For later use, we define
\begin{equation}
    \sigma^\sabc_{\rm L} = \frac{1}{2} \Re\left (\sigma^\sabc + \sigma^{s,a;cb} \right)
\end{equation}
and
\begin{equation}
    \sigma^\sabc_{\rm C} = \frac{1}{2} \Im\left (\sigma^\sabc - \sigma^{s,a;cb}\right)\,.
\end{equation}

The symmetry of the charge shift and injection conductivities were studied in Ref.~\cite{2020AhnBPVE}.
Here, we complete the analysis by studying the symmetry of the Drude, BCD, and rectification conductivities as well as the spin photoconductivities.
Let us begin with the charge case.

Following Ref.~\cite{2020AhnBPVE}, we first study how the matrix elements transform under a point-group symmetry operation and the time-reversal operation.
We do not assume that the system is invariant under the symmetry operations.
Instead, we study the relationship between the matrix element and conductivity tensors of the transformed and the original systems.
We denote the quantities of the transformed system with a prime.

First, let us consider a point-group symmetry operation $\opmcM$  whose real $3\times3$ transformation matrix is $\mcM$.
We denote the Hamiltonian of the original and transformed systems by $H_0$ and $H_0'=\opmcM\, H_0\,\opmcM^{-1}$, respectively. Also, $H_0({\bf k})=\exp(-i{\bf k}\cdot{\bf r})\,H_0\,\exp(i{\bf k}\cdot{\bf r})$ and $H_0'({\bf k})=\exp(-i{\bf k}\cdot{\bf r})\,H_0'\,\exp(i{\bf k}\cdot{\bf r})$.
The periodic parts of the Bloch states of the two systems satisfy
$\ket{u'_\nk} = \opmcM \ket{u_{n\mcM^{-1}\mb{k}}}$.
The eigenvalues satisfy $\veps'_\nk = \veps_{n\mcM^{-1}\mb{k}}$.
The velocity matrix element transforms as a vector:
\begin{widetext}
\begin{align} \label{eq:supp_symm_v_M}
    v_{mn}^{\prime a}(\mb{k})
    =& \hbar^{-1} \mel{u'_\mk}{\der{a}H_0'(\mb{k})}{u'_\nk} \nnnl
    =& \hbar^{-1} \mel{\opmcM u_{m\mcM^{-1}\mb{k}}} {\der{a}\left[\exp(-i{\bf k}\cdot{\bf r})\,\opmcM\,H_0\,\opmcM^{-1}\,\exp(i{\bf k}\cdot{\bf r})\,\right]} {\opmcM u_{n\mcM^{-1}\mb{k}}} \nnnl
    =& \hbar^{-1} \mel{\opmcM u_{m\mcM^{-1}\mb{k}}} {\opmcM\,\der{a}\left[\exp(-i{\bf k}\cdot\mcM{\bf r})\,H_0\,\exp(i{\bf k}\cdot\mcM{\bf r})\,\right]\,\opmcM^{-1}} {\opmcM u_{n\mcM^{-1}\mb{k}}} \nnnl
    =& \hbar^{-1} \mel{\opmcM u_{m\mcM^{-1}\mb{k}}} {\opmcM\,\der{a}\left[H_0(\mcM^{-1}\mb{k})\right]\,\opmcM^{-1}} {\opmcM u_{n\mcM^{-1}\mb{k}}} \nnnl
    =& \frac{\partial (\mcM^{-1}\mb{k})_b}{\partial k_a}\,\hbar^{-1} \mel{u_{m\mcM^{-1}\mb{k}}} {(\der{b}H_0)(\mcM^{-1}\mb{k})} {u_{n\mcM^{-1}\mb{k}}} \nnnl
    =& {\mcM^{-1}}_{ba}\, v^b_{mn}\left(\mcM^{-1}\mb{k}\right)\nnnl
    =& \mcM_{ab}\, v^b_{mn}(\mcM^{-1}\mb{k})\,.
\end{align}
The derivative operation also transforms as a vector:
\begin{equation} \label{eq:supp_symm_df_M}
    (\der{a} f)'(\mb{k})
    = \der{a} f(\mcM^{-1}\mb{k})
    = \frac{\partial (\mcM^{-1}\mb{k})_b}{\partial k_a} (\der{b} f)(\mcM^{-1}\mb{k})
    = \mcM_{ab} (\der{b} f)(\mcM^{-1}\mb{k}).
\end{equation}
\end{widetext}
Similarly, one can easily show that the generalized derivative $d^{0,b;a}_{mn}$ transforms as a rank-2 tensor,
\begin{align} \label{eq:supp_symm_d0_M}
    d^{\,\prime\, 0,b';a'}_{mn}(\mb{k})
    = \mcM_{a'a} \mcM_{b'b} d^{0,b;a}_{mn}(\mcM^{-1}\mb{k})\,,
\end{align}
and that
each of all the charge photoconductivity tensors (Drude, BCD, shift, injection, and rectification conductivity tensors) transforms under $\opmcM$ as
\begin{equation}
    \sigma^{\,\prime\, 0,a';b'c'} = \mcM_{a'a} \mcM_{b'b} \mcM_{c'c} \sigma^{0,\abc}\,.
\end{equation}
Thus, the second-order charge photoconductivity $\sigma^{0,a;bc}$ is a rank-3 tensor.

Next, let us consider the time-reversal operation $\opmcT$.
The periodic parts of the Bloch states of the transformed and the original systems satisfy
\begin{equation}
    \ket{u'_\nk} = \opmcT \ket{u_{n-\mb{k}}}.
\end{equation}
The velocity matrix element transforms as
\begin{align} \label{eq:supp_symm_v_T}
    v_{mn}^{\prime a}(\mb{k})
    =& \hbar^{-1} \mel{u'_\mk} {\der{a}H_0'(\mb{k})} {u'_\nk} \nnnl
    =& \hbar^{-1} \mel{\opmcT u_{m-\mb{k}}} {\der{a} \left[\opmcT H_0(-\mb{k}) \opmcT^{-1} \right]} {\opmcT u_{n-\mb{k}}} \nnnl
    =& \hbar^{-1} \braket{\opmcT u_{m-\mb{k}}} {\opmcT \der{a} \left[H_0(-\mb{k}) \right] u_{n-\mb{k}}} \nnnl
    =& \hbar^{-1} \braket{\der{a} \left[H_0(-\mb{k}) \right] u_{n-\mb{k}}} {u_{m-\mb{k}}} \nnnl
    =& - \hbar^{-1} \braket{ (\der{a} H_0)(-\mb{k}) u_{n-\mb{k}}} {u_{m-\mb{k}}} \nnnl
    =& - v_{nm}^a (-\mb{k})
\end{align}
In the fourth equality, we used $\braket{\opmcT u}{\opmcT v} = \braket{v}{u}$.
The derivative operation also transforms as
\begin{equation} \label{eq:supp_symm_df_T}
    (\der{a} f)'(\mb{k})
    = \der{a} f(-\mb{k})
    = -(\der{a} f)(-\mb{k})\,.
\end{equation}
The generalized derivative $d^{0,b;a}_{mn}$ transforms as
\begin{alignat}{2} \label{eq:supp_symm_d0_T}
    d^{\,\prime\, 0,b;a}_{mn}(\mb{k})
    =& j^{0,ab}_{nm}(-\mb{k})
    &&+ \sum_{p\neq m} \frac{v^a_{pm}(-\mb{k}) v^b_{np}(-\mb{k})}{\omega_{mp}(-\mb{k})} \nnnl
    & &&+ \sum_{p\neq n} \frac{v^b_{pm}(-\mb{k}) v^a_{np}(-\mb{k})}{\omega_{np}(-\mb{k})} \nnnl
    =& d^{0,b;a}_{nm}(-\mb{k}).
\end{alignat}

Now, using the symmetry properties of the matrix elements, we study the transformation of charge conductivity tensors under $\opmcT$.

\begin{widetext}
\subsection{Drude} \label{sec:supp_symmetry_drude}
Equation~\eqref{eq:sigma_drude} shows that the charge Drude conductivity is always symmetric in the exchange of $b$ and $c$.
Hence, we find
\begin{equation}
    \sigma^{0,\abc}_{\rm Drude, C} = \frac{1}{2} \Im \left(\sigma^{0,\abc}_{\rm Drude} - \sigma^{0,a;cb}_{\rm Drude} \right) = 0.
\end{equation}
Further from \myeqref{eq:sigma_drude}, one can easily show that the charge Drude conductivity transforms as
\begin{equation}
    \sigma^{\,\prime\, 0,\abc}_\mathrm{Drude}
    = - \sigma^{0,\abc}_\mathrm{Drude}.
\end{equation}
The $(-1)$ factor shows that the charge Drude conductivity is odd under $\opmcT$.
Thus, in a $\opmcT$-symmetric system, the charge Drude conductivity is zero because
$\sigma^{0,a;bc}_\mathrm{Drude} = \sigma^{\,\prime\, 0,a;bc}_\mathrm{Drude} = - \sigma^{0,a;bc}_\mathrm{Drude}$ holds.

One can use $j^{0,a}_{mm} = v^a_{mm} = \der{a}\veps_m / \hbar$ and partial integration to rewrite the charge Drude conductivity as follows:
\begin{align} \label{eq:supp_symm_drude_3}
    \sigma^{0,\abc}_\mathrm{Drude}
    = \frac{q^3}{2\hbar^2 V \Omega^2} \sum_{\mb{k},m} j^{0,a}_{mm} \der{b} \der{c} f_m
    = \frac{q^3}{2\hbar^3 V \Omega^2} \sum_{\mb{k},m} (\der{a} \veps_m) (\der{b} \der{c} f_m)
    = \frac{q^3}{2\hbar^3 V \Omega^2} \sum_{\mb{k},m} (\der{a}\der{b}\der{c} \veps_m) f_m.
\end{align}
Hence, the charge Drude current is symmetric under the permutation of all three indices.

\subsection{Berry curvature dipole}
For the BCD current, \myeqref{eq:sigma_BCD} shows that it is always anti-symmetric in the exchange of $b$ and $c$.
Hence, we find
\begin{equation}
    \sigma^{0,\abc}_{\rm BCD, L} = \frac{1}{2} \Re \left(\sigma^{0,\abc}_{\rm BCD} + \sigma^{0,a;cb}_{\rm BCD} \right) = 0.
\end{equation}
The summand of \myeqref{eq:sigma_BCD} transforms under $\opmcT$ as
\begin{equation} \label{eq:supp_summ_BCD_mel}
    \left[ \frac{\Im(j_{mn}^{0,a} v_{nm}^{b})}{\omega_{mn}^2} (\der{c} f_m) \right]' (\mb{k})
    = \left[ \frac{\Im(j_{nm}^{0,a} v_{mn}^{b})}{\omega_{mn}^2} (-\der{c} f_m) \right] (-\mb{k})
    = \left[ \frac{\Im(j_{mn}^{0,a} v_{nm}^{b})}{\omega_{mn}^2} (\der{c} f_m) \right] (-\mb{k})
\end{equation}
Hereafter, we use the notation that the arguments $\mb{k}$ and $-\mb{k}$ as well as the prime apply to all matrix elements inside the outermost parentheses on its left.
In the last equality of \myeqref{eq:supp_summ_BCD_mel}, we used the Hermiticity of the current operators.
Thus, the charge BCD conductivity is symmetric under $\opmcT$:
\begin{equation}
    \sigma^{\,\prime\, 0,\abc}_\mathrm{BCD}
    = \sigma^{0,\abc}_\mathrm{BCD}.
\end{equation}

\subsection{Shift}
Next, let us consider the shift conductivity.
Using Eqs.~(\ref{eq:supp_symm_v_T}, \ref{eq:supp_symm_d0_T}), we find that the matrix element in \myeqref{eq:sigma_shift} transforms as
\begin{equation}
    \left( d^{0,b;a}_{mn} v^c_{nm} - d^{0,c;a}_{nm} v^b_{mn} \right)' (\mb{k})
    = -\left( d^{0,b;a}_{nm} v^c_{mn} - d^{0,c;a}_{mn} v^b_{nm} \right) (-\mb{k})
    = \left( d^{0,c;a}_{mn} v^b_{nm} - d^{0,b;a}_{nm} v^c_{mn} \right) (-\mb{k}).
\end{equation}
Note that the time reversal exchanges the field direction indices $b$ and $c$ without a sign flip.
Thus, the charge shift conductivity transforms under $\opmcT$ as
\begin{equation}
    \sigma^{\,\prime\, 0,\acb}_\mathrm{shift}
    = \sigma^{0,\abc}_\mathrm{shift}.
\end{equation}
The shift conductivity for linearly polarized light transforms under $\opmcT$ as
\begin{equation} \label{eq:supp_symm_shift_L}
    \sigma^{\,\prime\, 0,\abc}_\mathrm{shift, L}
    = \frac{1}{2} \Re \left( \sigma^{\,\prime\, 0,\abc}_\mathrm{shift} + \sigma^{\,\prime\, 0,\acb}_\mathrm{shift} \right)
    = \frac{1}{2} \Re \left( \sigma^{0,\acb}_\mathrm{shift} + \sigma^{0,\abc}_\mathrm{shift} \right)
    = \sigma^{0,\abc}_\mathrm{shift, L},
\end{equation}
while that for circularly polarized light transforms as
\begin{equation} \label{eq:supp_symm_shift_C}
    \sigma^{\,\prime\, 0,\abc}_\mathrm{shift, C}
    = \frac{1}{2} \Im \left( \sigma^{\,\prime\, 0,\abc}_\mathrm{shift} - \sigma^{\,\prime\, 0,\acb}_\mathrm{shift} \right)
    = \frac{1}{2} \Im \left( \sigma^{0,\acb}_\mathrm{shift} - \sigma^{0,\abc}_\mathrm{shift} \right)
    = - \sigma^{0,\abc}_\mathrm{shift, C}.
\end{equation}
The shift conductivity for linearly (circularly) polarized light is even (odd) under time reversal.

\subsection{Injection}
Next, let us consider the injection conductivity.
Using Eqs.~(\ref{eq:supp_symm_v_T}), we find that the matrix elements in \myeqref{eq:sigma_inj} transforms under $\opmcT$ as
\begin{equation}
    \left( f_{mn} \Delta j^{0,a}_{mn} v^b_{mn} v^c_{nm} \right)' (\mb{k}) =
    - \left( f_{mn} \Delta j^{0,a}_{mn} v^b_{nm} v^c_{mn} \right) (-\mb{k})
\end{equation}
The time reversal exchanges the field direction indices $b$ and $c$ and gives a sign flip.
Thus, the rectification conductivity transforms under $\opmcT$ as
\begin{equation} \label{eq:supp_symm_inj}
    \sigma^{\,\prime\, 0,\acb}_\mathrm{inj.} = -\sigma^{0,\abc}_\mathrm{inj.}.
\end{equation}
The injection conductivity for linearly polarized light transforms under $\opmcT$ as
\begin{equation} \label{eq:supp_symm_inj_L}
    \sigma^{\,\prime\, 0,\abc}_\mathrm{inj., L}
    = \frac{1}{2} \Re \left( \sigma^{\,\prime\, 0,\abc}_\mathrm{inj.} + \sigma^{\,\prime\, 0,\acb}_\mathrm{inj.} \right)
    = -\frac{1}{2} \Re \left( \sigma^{0,\acb}_\mathrm{inj.} + \sigma^{0,\abc}_\mathrm{inj.} \right)
    = -\sigma^{0,\abc}_\mathrm{inj., L},
\end{equation}
while that for circularly polarized light transforms as
\begin{equation} \label{eq:supp_symm_inj_C}
    \sigma^{\,\prime\, 0,\abc}_\mathrm{inj., C}
    = \frac{1}{2} \Im \left( \sigma^{\,\prime\, 0,\abc}_\mathrm{inj.} - \sigma^{\,\prime\, 0,\acb}_\mathrm{inj.} \right)
    = -\frac{1}{2} \Im \left( \sigma^{0,\acb}_\mathrm{inj.} - \sigma^{0,\abc}_\mathrm{inj.} \right)
    = \sigma^{0,\abc}_\mathrm{inj., C}.
\end{equation}
Note the sign difference between Eqs.~(\ref{eq:supp_symm_shift_L}, \ref{eq:supp_symm_shift_C}) and Eqs.~(\ref{eq:supp_symm_inj_L}, \ref{eq:supp_symm_inj_C}).
The injection conductivity for linearly (circularly) polarized light is odd (even) under time reversal.

\subsection{Rectification}
Finally, let us consider the rectification conductivity.
Using Eqs.~(\ref{eq:supp_symm_v_T}, \ref{eq:supp_symm_d0_T}), one can show the term in the inner parentheses of \myeqref{eq:sigma_rec} transforms under $\opmcT$ as
\begin{equation}
    \left( d^{0,b;a}_{mn} v^c_{nm} + d^{0,c;a}_{nm} v^b_{mn}
    - \frac{2\Delta j^{0,a}_{mn} v^b_{mn} v^c_{nm}}{\omega_{mn}} \right)' (\mb{k})
    =
    -\left( d^{0,b;a}_{nm} v^c_{mn} + d^{0,c;a}_{mn} v^b_{nm}
    - \frac{2\Delta j^{0,a}_{mn} v^b_{nm} v^c_{mn}}{\omega_{mn}} \right) (-\mb{k}).
\end{equation}
The time reversal exchanges the field direction indices $b$ and $c$ and gives a sign flip.
Thus, the rectification conductivity transforms under $\opmcT$ as
\begin{equation}
    \sigma^{\,\prime\, 0,\acb}_\mathrm{rect.} = -\sigma^{0,\abc}_\mathrm{rect.},
\end{equation}
which has the same form as \myeqref{eq:supp_symm_inj}.
Thus, the symmetry property of the rectification conductivity is identical to that of the injection conductivity.
This result completes the symmetry analysis of second-order charge photocurrents.

\subsection{Spin shift vector}
Now, let us discuss the symmetry transformation property of the spin shift vector, which we have proposed as a key quantity that describes the spin shift and rectification currents.
Note that in the shift vector $\Rtilde^{s,\mcE;a}_{mn}$, one should not treat the light polarization $\mb{\mcE}$ as a tensor index because the shift vector does not transform like a tensor for that index (see the second term of \myeqref{eq:spin_shift_vector}).
Hence, we treat $\mb{\mcE}$ as an external parameter that is transformed together with the system.

First, let us consider a rotation operation $\hat{\mcM}$, whose real $3\times3$ transformation matrix is $\mcM$.
The light polarization vector transforms as $\mcE^{'a} = \mcM_{ab} \mcE^b$.
If an operator $\mathcal{O}^a(\mb{k})$ transforms as $\mathcal{O}'^{a}_{mn}(\mb{k}) = \mcM_{ab} \mathcal{O}^{b}_{mn}(\mcM^{-1}\mb{k})$ under a rotation $\mcM$, we find that $\mathcal{O}^{\mcE}_{mn}$ transforms as
\begin{equation} \label{eq:supp_symm_O_E_M}
    \left(\mathcal{O}^{\mcE}_{mn}\right)'(\mb{k})
    = \mathcal{O}'^a_{mn}(\mb{k}) \mcE'^a
    = \mcM_{ab} \mcM_{ac} \mathcal{O}^b_{mn}(\mcM^{-1}\mb{k}) \mcE^c
    = \delta_{b,c} \mathcal{O}^b_{mn}(\mcM^{-1}\mb{k}) \mcE^c
    = \mathcal{O}^{\mcE}_{mn}(\mcM^{-1}\mb{k}).
\end{equation}
In other words, the inner product of two vector quantities [$\mathcal{O}^a_{mn}(\mb{k})$ and $\mcE^a$] transforms as a scalar.
Then, one finds that the spin shift vector~[\myeqref{eq:spin_shift_vector}] transforms as
\begin{equation} \label{eq:supp_symm_shift_vector_M}
    \left(\Rtilde^{s',\mcE;a'}_{mn}\right)'(\mb{k})
    = i\left( \frac{d^{s',\mcE;a'}_{mn}}{v^{\mcE}_{mn}} - \frac{\Delta j^{s',a'}_{mn}}{\omega_{mn}} \right)'(\mb{k})
    = i \mcM_{s's} \mcM_{a'a} \left( \frac{d^{s,\mcE;a}_{mn}}{v^\mcE_{mn}} - \frac{\Delta j^{s,a}_{mn}}{\omega_{mn}} \right)(\mcM^{-1}\mb{k})
    = \mcM_{s's} \mcM_{a'a} \Rtilde^{s,\mcE;a}_{mn}(\mcM^{-1}\mb{k}).
\end{equation}
For an improper rotation, the spin shift vector gets an additional $-1$ factor because the spin-current operator is an axial vector.

Next, for the time-reversal operation, the light polarization transforms as $\mcE'^{a} = (\mcE^a)^*$.
If $\mathcal{O}^a$ transforms under time reversal as $\mathcal{O}'^{a}_{mn}(\mb{k}) = \pm\left[\mathcal{O}^{a}_{mn}(-\mb{k})\right]^*$, $\mathcal{O}^\mcE$ transforms as
\begin{equation} \label{eq:supp_symm_O_E_T}
    \left(\mathcal{O}^{\mcE}_{mn}\right)'(\mb{k})
    = \mathcal{O}'^a_{mn}(\mb{k}) \mcE'^a
    = \pm\left[\mathcal{O}^{a}_{mn}(-\mb{k}) \right]^* (\mcE^a)^*
    = \pm\left[\mathcal{O}^{\mcE}_{mn}(-\mb{k}) \right]^*.
\end{equation}
Then, for $s\neq0$ we find
\begin{equation} \label{eq:supp_symm_shift_vector_T}
    \left(\Rtilde^{s,\mcE;a}_{mn}\right)'(\mb{k})
    = \left[ i\left( \frac{d^{s,\mcE;a}_{mn}}{v^{\mcE}_{mn}} - \frac{\Delta j^{s,a}_{mn}}{\omega_{mn}} \right) \right]'(\mb{k})
    = i \left[\left(\frac{d^{s,\mcE;a}_{mn}}{v^{\mcE}_{mn}} - \frac{\Delta j^{s,a}_{mn}}{\omega_{mn}} \right) (-\mb{k})\right]^*
    = -\left(\Rtilde^{s,\mcE;a}_{mn}(-\mb{k})\right)^*.
\end{equation}
In the second equality, we included the $-1$ factor that occurs because the spin-current operator is odd under time reversal.
For the charge case, one finds
\begin{equation} \label{eq:supp_symm_shift_vector_T_0}
    \left(\Rtilde^{0,\mcE^;a}_{mn}\right)'(\mb{k})
    = \left(\Rtilde^{0,\mcE;a}_{mn}(-\mb{k})\right)^*.
\end{equation}
\end{widetext}

% =========================================================
\section{Computataional details}
\subsection{Details of density functional theory calculations}
We used the \qe\ package~\cite{2017GiannozziQE} to perform density functional theory calculations.
For the self-consistent field calculation, we used an unshifted 16$\times$16$\times$12 $k$-point grid, a kinetic energy cutoff of 80~Ry, 
fully relativistic ONCV pseudopotentials~\cite{2013HamannONCVPSP} taken from the PseudoDojo library (v0.4)~\cite{2018VanSettenPseudoDojo}, and the Perdew-Burke-Ernzerhof (PBE) exchange-correlation functional~\cite{1996PerdewPBE}.
The lattice parameters and atomic coordinates were optimized until the stresses and forces were less than $1.0 \times 10^{-6}$~Ry/Bohr$^3$ and $6.0 \times 10^{-4}$~Ry/Bohr, respectively.
The optimized lattice parameters were $a$=4.44~\AA\ and $c$=7.39~\AA, consistent with those of Ref.~\cite{2020BrousseauBiTeI}.
Although the PBE functional overestimates lattice parameters compared to the experimental values~\cite{2011IshizakaBiTeI}, it gives band gap and Rashba splitting in reasonable agreement with experiment~\cite{2020BrousseauBiTeI}.

We used the \wannier\ package~\cite{2020PizziWannier90} to construct the Wannier-function-based tight-binding model.
The Brillouin zone was sampled with an 8$\times$8$\times$6 grid for Wannierization.
We construct 18 Wannier functions using the atomic $p$ orbitals as initial guesses.
Disentanglement was not used since the target bands are isolated from other bands.
To preserve the crystal symmetries, we did not perform the maximal localization step.
The centers of the Wannier functions for the tight-binding model were calculated using the translationally invariant formula: Eq.~(31) of Ref.~\cite{1997MarzariMLWF}.

\subsection{Details of photoconductivity calculations}
We used a modified version of \wannier~\cite{2020PizziWannier90} for the Wannier-interpolation calculation of the photoconductivity.
We sampled the Brillouin zone using a grid shifted by half the grid spacing along all axes.
Using the shifted grid speeded up the convergence.
The photoconductivity was calculated using a dense $k$-point grid of 800$\times$800$\times$800 to obtain converged values.
A fixed numerical smearing of 20~meV was applied to the delta functions and principal values involving the energy difference between the initial and final states.
A detailed convergence study is shown in Figs.~\ref{fig:supp_converge_sigma_nk}-\ref{fig:supp_converge_rect}.
The Fermi-Dirac occupation was calculated at zero temperature.

To avoid numerical problems related to near-degenerate states, we regularized the denominator including intermediate states, such as $1/\omega_{np}$ in \myeqref{eq:sder_def}, to $\Re[1/(\omega_{np} + i \eta)]$~\cite{2006Nastos,2018IbanezShift}.
The broadening parameter was set to $\eta = 1~\mathrm{meV}$ unless noted otherwise.

We use the diagonal tight-binding approximation (TBA) within which the position matrix elements between different Wannier functions are neglected~\cite{2018IbanezShift}.
This approximation is needed because using the existing Wannier interpolation methods, it is theoretically impossible to calculate the spin-velocity derivative [\myeqref{eq:sder_def}] without erroneously truncating the summation over bands.
By using the diagonal tight-binding approximation, the sum rules are modified to include only a finite number of bands~\cite{2018IbanezShift} and are satisfied without any truncation error.

The accuracy of the diagonal tight-binding approximation can be tested by calculating the charge shift current and charge and spin injection currents without the approximation.
Such a calculation is possible because the injection current [\myeqref{eq:sigma_inj}] does not involve the spin-velocity derivative, and a truncation-error-free expression exists for the charge shift current~\cite{2018IbanezShift}.
In Figs.~\ref{fig:supp_w90_shift} and \ref{fig:supp_w90_inj}, we show that the diagonal tight-binding approximation changes the photoconductivity only slightly.

%========================================================
\section{Additional computational results}
\label{sec:supp_additional}

\begin{figure*}[htb]
\centering
\includegraphics[width=0.8\textwidth]{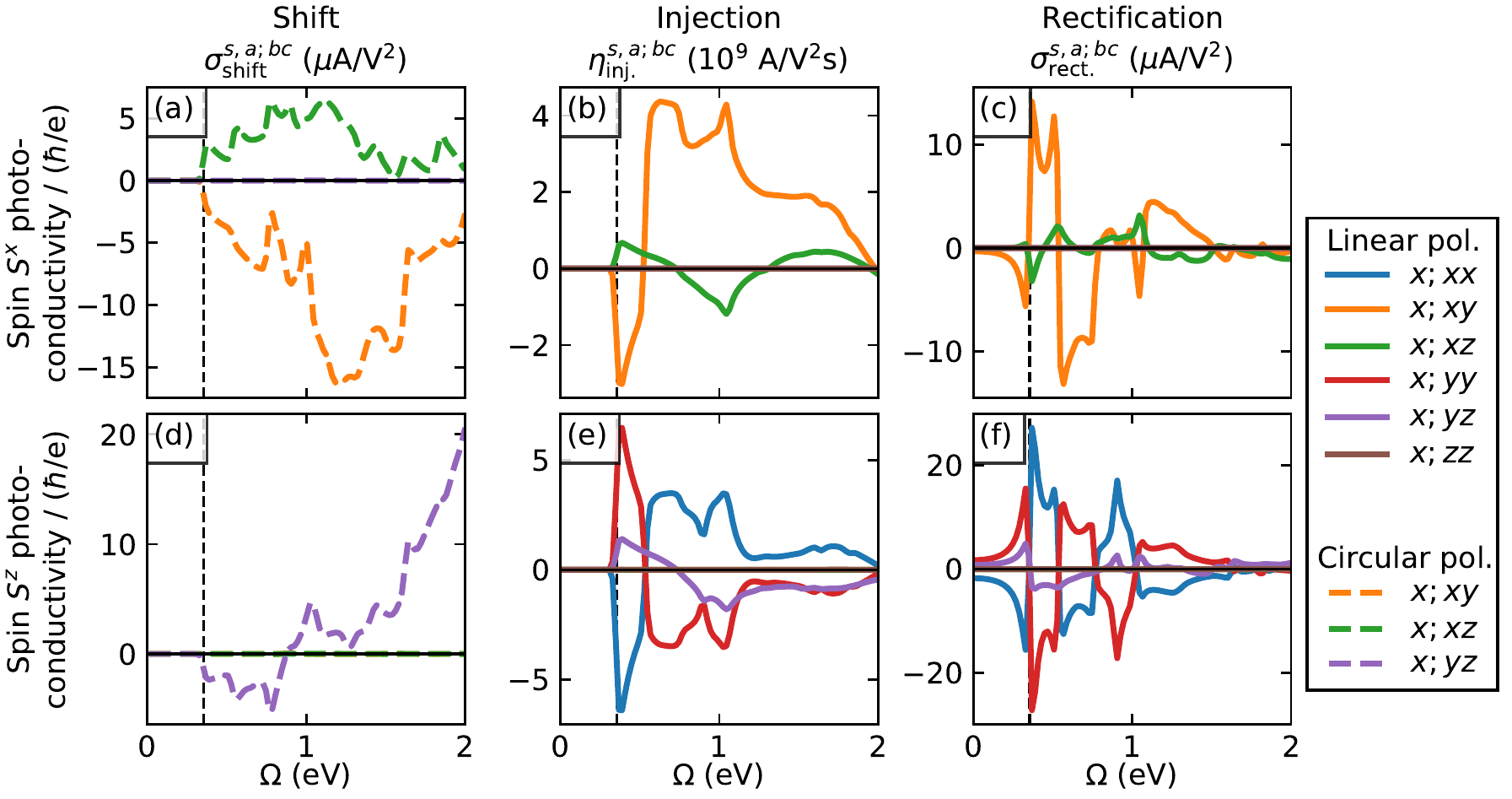}
\caption{
(a-c) Spin $S^x$ and (d-f) spin $S^z$ photoconductivity of BiTeI.
The vertical dashed lines indicate the direct band gap of 0.35~eV.
}
\label{fig:supp_sigma_xz}
\end{figure*}

\begin{figure}[htb]
\centering
\includegraphics[width=1.0\columnwidth]{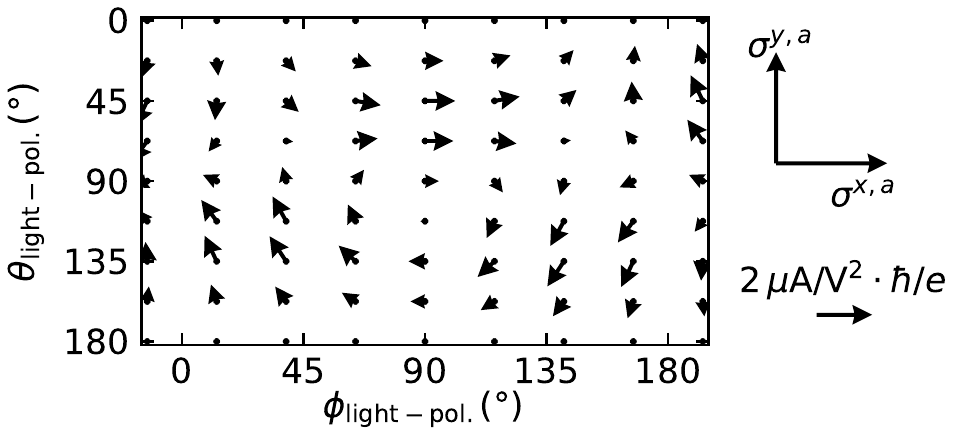}
\caption{
Spin polarization of the subgap pure spin current along the $z$ direction.
The arrows indicate the in-plane spin polarization of the current.
The $S^z$ component is zero due to symmetry.
Light frequency is $\Omega=0.30$~eV.
The scale of the arrow is smaller than in Fig.~\ref{fig:subgap_pol}.
}
\label{fig:supp_subgap_pol_z}
\end{figure}

\begin{figure*}[htb]
\centering
\includegraphics[width=1.0\textwidth]{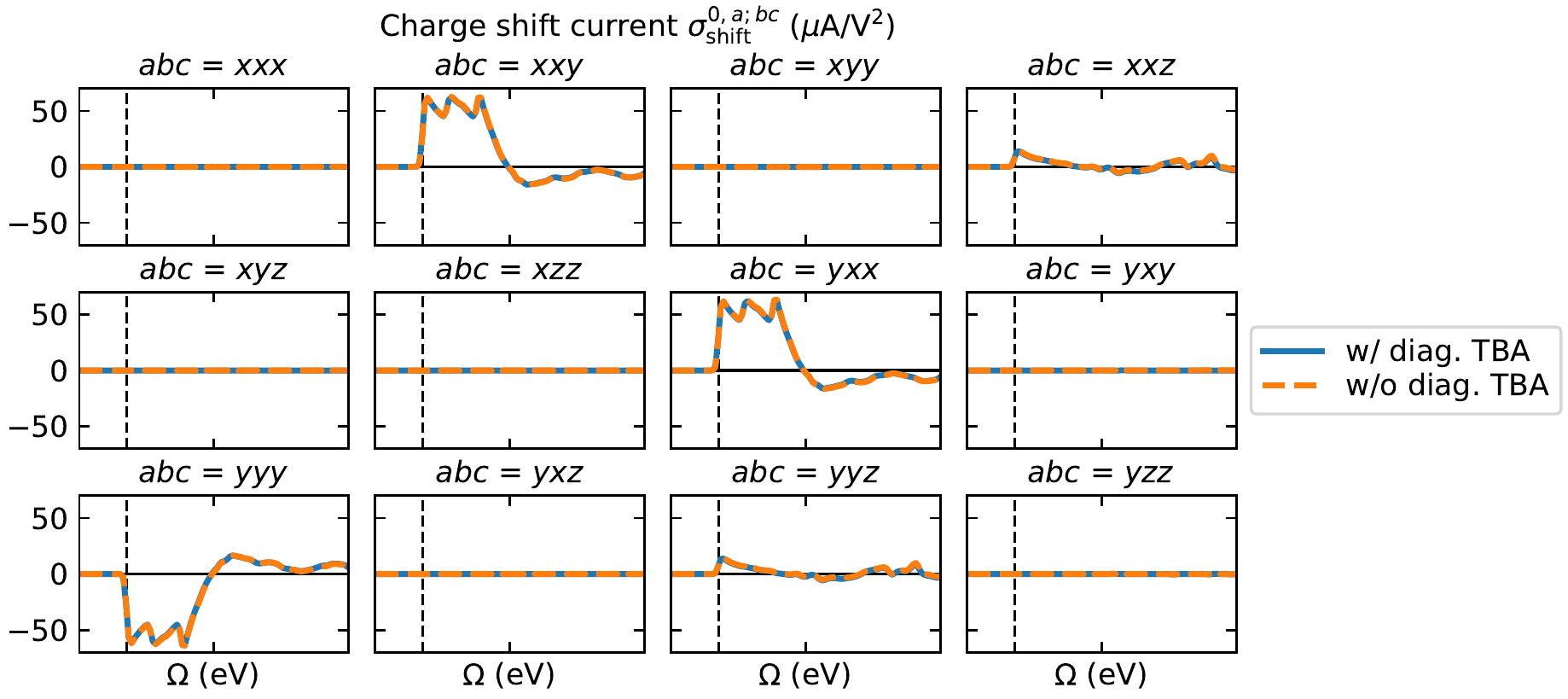}
\caption{
Charge shift current calculated with and without the diagonal tight-binding approximation (diag. TBA).
}
\label{fig:supp_w90_shift}
\end{figure*}

\begin{figure*}[htb]
\centering
\includegraphics[width=1.0\textwidth]{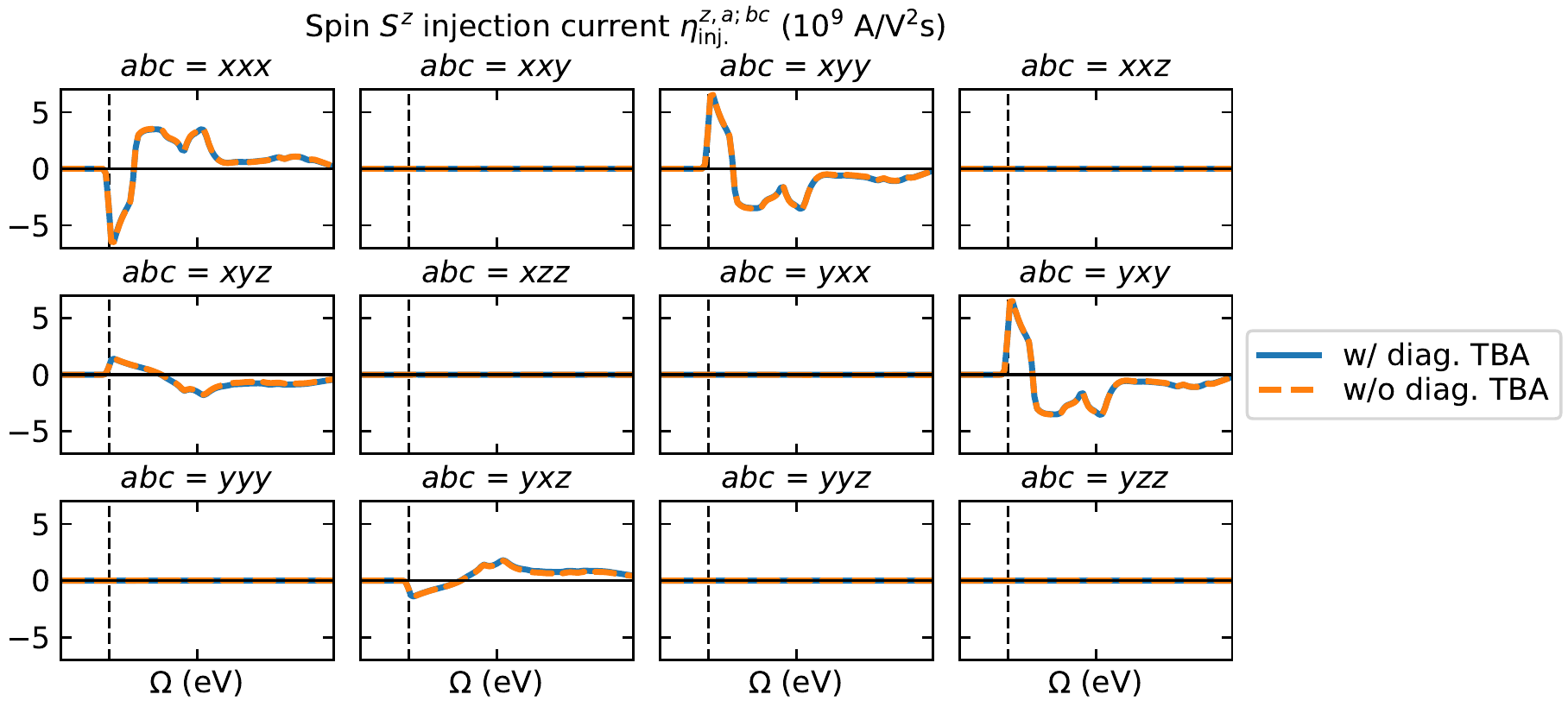}
\caption{
Spin $S^z$ injection current calculated with and without the diagonal tight-binding approximation.
}
\label{fig:supp_w90_inj}
\end{figure*}

\begin{figure*}[htb]
\centering
\includegraphics[width=0.7\textwidth]{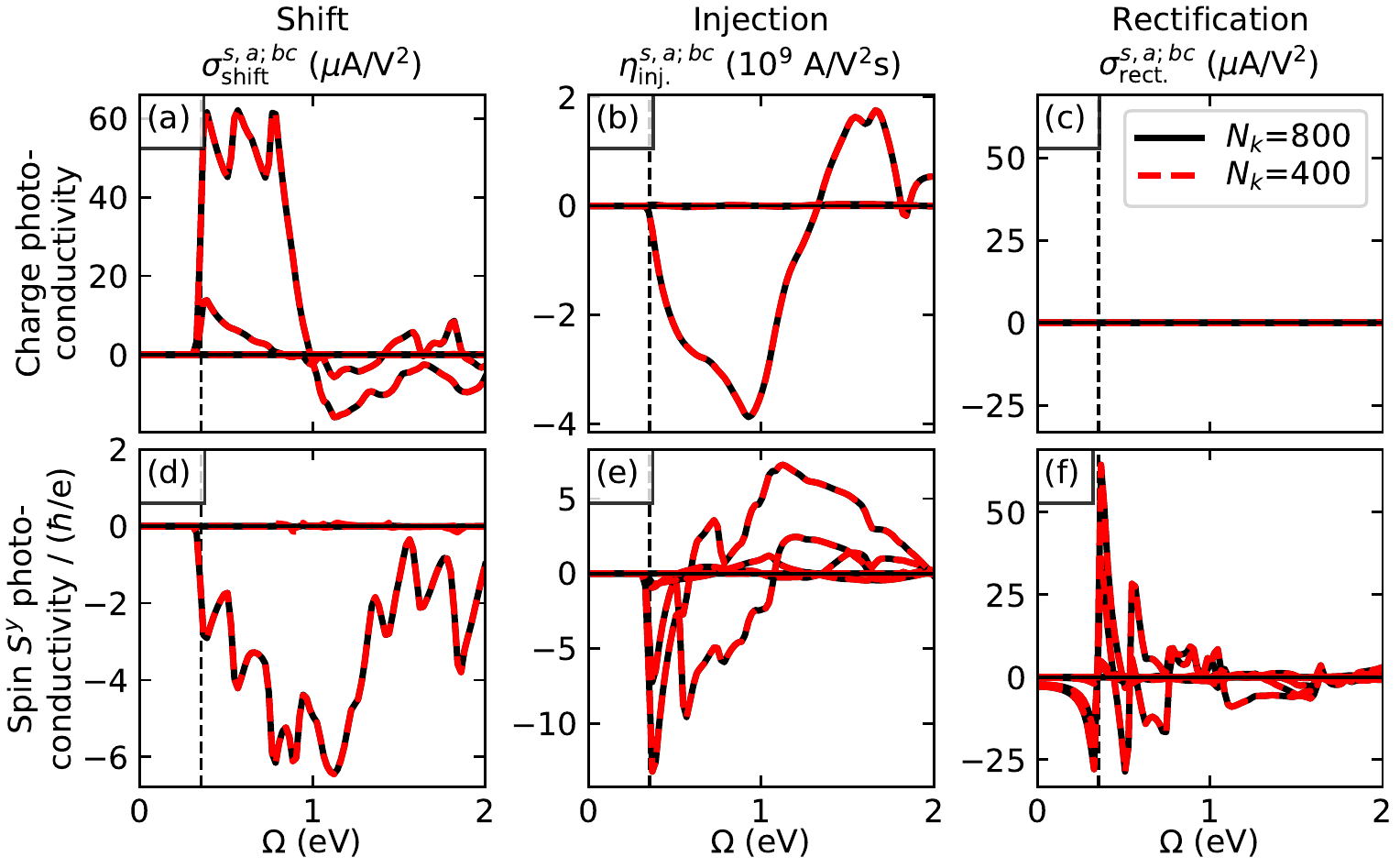}
\caption{
Convergence of photoconductivity with respect to the size of the $k$-point grid.
(a-c) Charge and (d-f) spin $S^z$ photoconductivity calculated with an $N_k\times N_k\times N_k$ $k$-point grid.
The plotted photoconductivity tensor components are those plotted in Fig.~\ref{fig:sigma_all}(B-G).
}
\label{fig:supp_converge_sigma_nk}
\end{figure*}

\begin{figure*}[htb]
\centering
\includegraphics[width=0.7\textwidth]{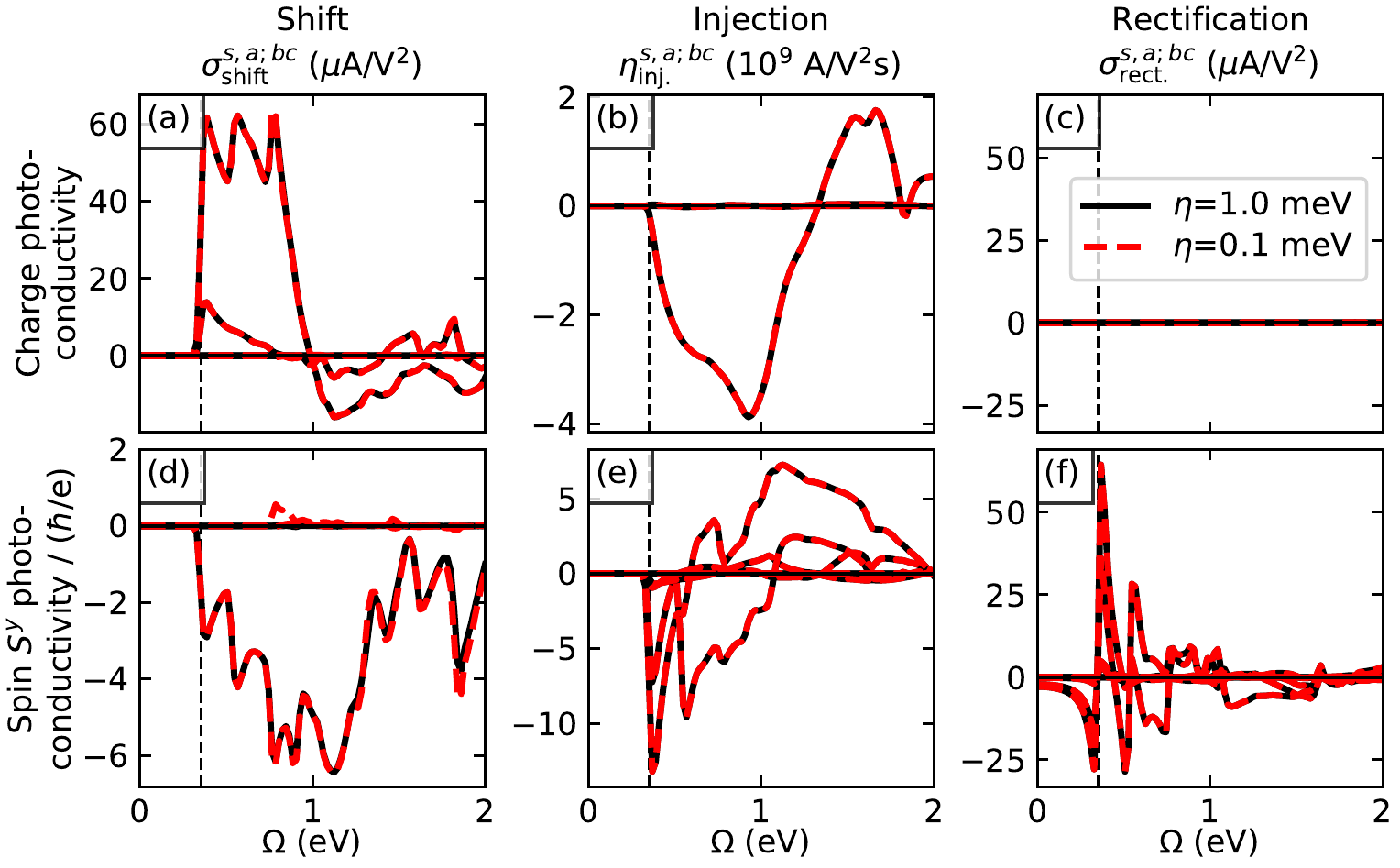}
\caption{
Convergence of photoconductivity with respect to $\eta$.
(a-c) Charge and (d-f) spin $S^z$ photoconductivity calculated with $\eta=$1~meV and 0.1~meV.
The plotted photoconductivity tensor components are those plotted in Fig.~\ref{fig:sigma_all}(B-G).
}
\label{fig:supp_converge_sigma_eta}
\end{figure*}

\begin{figure}[htb]
\centering
\includegraphics[width=1.0\columnwidth]{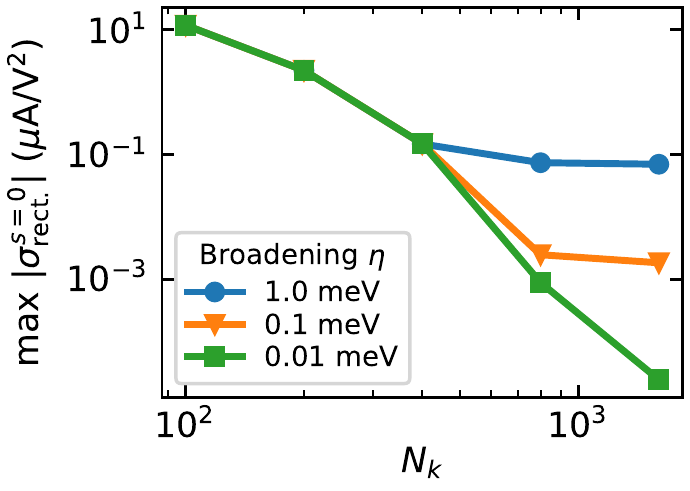}
\caption{
Convergence of charge rectification conductivity.
The maximum absolute value is calculated in the 0~eV to 2~eV window.
The size of the $k$-point grid is $N_k \times N_k \times N_k$.
The lines are a guide to the eye.
}
\label{fig:supp_converge_rect}
\end{figure}

Figure~\ref{fig:supp_sigma_xz} shows the spin $S^x$ and $S^z$ photoconductivity for currents flowing along the $x$ direction.

Figure~\ref{fig:supp_subgap_pol_z} shows the spin polarization of the subgap spin conductivity for currents flowing along the $z$ direction as a function of light polarization angles.
Note that the size of the spin conductivity shown in Fig.~\ref{fig:supp_subgap_pol_z} is an order of magnitude smaller than that of Fig.~\ref{fig:subgap_pol}.
The spin photocurrent flowing along the $z$ direction in BiTeI is small because the corresponding group velocity of the carriers is small in BiTeI, a layered compound.

In Fig.~\ref{fig:supp_w90_shift} and Fig.~\ref{fig:supp_w90_inj}, we show the charge shift and spin injection currents calculated with and without the diagonal tight-binding approximation.
We find that the diagonal tight-binding approximation changes the photocurrent only slightly.

Figures~\ref{fig:supp_converge_sigma_nk}-\ref{fig:supp_converge_rect} show the convergence of the photoconductivity with respect to the $k$-point grid size.
In Fig.~\ref{fig:supp_converge_sigma_nk}, we find that all photoconductivity elements are already converged at a $400\times400\times400$ $k$-point grid.
Figure~\ref{fig:supp_converge_sigma_eta} shows that all photoconductivity elements are converged with respect to $\eta$ at $\eta=1$~meV.
In Fig.~\ref{fig:supp_converge_rect}, we show that the charge rectification current converges to zero in the limit of an infinitely fine $k$-point grid as $\eta \rightarrow 0$, as expected in insulators.

\FloatBarrier % Forces bibliography to appear after all the figures.

\bibliography{main}

\end{document}